\documentclass[pre,superscriptaddress,twocolumn,floatfix]{revtex4-2}

\usepackage{microtype}
\usepackage[utf8]{inputenx}
\usepackage{physics}                            
\usepackage{bm}									
\usepackage{enumerate}							
\usepackage{amsthm}								
\usepackage{comment}							
\usepackage[caption=false]{subfig}
\usepackage{amssymb}							
\usepackage{t1enc}								
\usepackage{mathtools}							
\usepackage{tensor}								
\usepackage{wrapfig}							
\usepackage{listings}							
\usepackage{xcolor}								
\usepackage{hyperref}
\hypersetup{colorlinks=true}
\usepackage[capitalize]{cleveref}               
\usepackage{graphicx,color}
\usepackage{amsfonts,amsmath}
\usepackage{xstring}                            
\usepackage{lettrine}
\usepackage{algorithm}
\usepackage{algorithmic}

\newcommand{\eg}{{\em e.g.}}
\newcommand{\ie}{{\em i.e.}}

\newcommand{\bint}[0]{\beta_{int}}
\newcommand{\bnat}[0]{\beta_{nat}}
\newcommand{\hbint}[0]{\hat{\beta}_{int}}
\newcommand{\hbnat}[0]{\hat{\beta}_{nat}}
\newcommand{\mycomment}[1]{}

\begin{document}

\title{Evolutionary Dynamics of a Lattice Dimer: \\ a Toy Model for Stability vs. Affinity Trade-offs in Proteins}

\author{E. Loffredo}
\email{emanuele.loffredo@phys.ens.fr}
\affiliation{ 
Laboratory of Physics of the Ecole Normale Sup\'erieure, CNRS UMR 8023 and PSL Research,
Sorbonne Universit\'e, 24 rue Lhomond, 75231 Paris cedex 05, France
}
\author{E. Vesconi}
\altaffiliation{Currently at \textit{Planven Entrepreneur Ventures, Glockengasse 9, 8001, Zurich}}
\affiliation{ 
Laboratory of Physics of the Ecole Normale Sup\'erieure, CNRS UMR 8023 and PSL Research,
Sorbonne Universit\'e, 24 rue Lhomond, 75231 Paris cedex 05, France
}
\author{R. Razban}
\affiliation{Laufer Center for Physical and Quantitative Biology, Stony Brook University, Stony Brook, NY, USA}
\author{O. Peleg}
\affiliation{Computer Science Dept, University of Colorado Boulder, Boulder, CO 80309, USA}
\affiliation{Santa Fe Institute, Santa Fe, NM 87501, USA (External
Faculty) and Ecology \& Evolutionary Biology, Physics,
Applied Math Dept, University of Colorado Boulder,
Boulder, CO 80309, USA}
\author{E. Shakhnovich}
\affiliation{Department of Chemistry and Chemical Biology, Harvard University, Cambridge, MA 02138}
\author{S. Cocco}
\author{R. Monasson}
\affiliation{ 
Laboratory of Physics of the Ecole Normale Sup\'erieure, CNRS UMR 8023 and PSL Research,
Sorbonne Universit\'e, 24 rue Lhomond, 75231 Paris cedex 05, France
}

\begin{abstract}
Understanding how a stressor applied on a biological system shapes its evolution is key to achieving targeted evolutionary control. Here we present a toy model of two interacting lattice proteins to quantify the response to the selective pressure defined by the binding energy. We generate sequence data of proteins and study how the sequence and structural properties of dimers are affected by the applied selective pressure, both during the evolutionary process and in the stationary regime. In particular we show that internal contacts of native structures lose strength, while inter-structure contacts are strengthened due to the folding-binding competition. We discuss how dimerization is achieved through enhanced mutability on the interacting faces, and how the designability of each native structure changes upon introduction of the stressor.
\end{abstract}

\maketitle

\section{Introduction}
\label{par:intro}
To be functional proteins must satisfy a variety of constraints, related to their stability, activity, specificity etc ... Most proteins reach an adequately folded stable conformation to efficiently perform their tasks involving binding with other biomolecules,  such as DNA, peptidic ligand or other proteins. How these various evolutionary constraints combine to shape the evolutionary landscape of proteins is of great interest. This question is particularly crucial when additional stressors, e.g. an antibiotic on a bacterial population, are dynamically applied during evolution. To what extent organisms can accommodate new constraints, while still fulfilling constitutive ones, is an important issue.

In this work we introduce a minimal setting to address this question from a theoretical point of view. We consider lattice proteins (LP) \cite{lau1989lattice, shakhnovich1993engineering, li1996emergence,shakhnovich1990enumeration}, an exactly solvable model of amino acids chains constrained to fold on the sites of a cube. Despite their simplicity, LPs share many features with real proteins and have been proved to be a useful tool for studying protein folding and designability. In addition to require that proteins acquire their native folds with high probability, we impose that they bind each other and form a stable dimer \cite{peleg2014evolution}. The study of the conflict between these two requirements, and its consequence on the distribution of adequate sequences is the goal of the present work.

In a first part of the work we study the stationary regime in which sequences evolve through a mutational dynamics under a two-fold selection pressure requiring the native folds and the dimer conformation to be achieved. We make use of the so-called direct coupling approximation (DCA) \cite{weigt2009identification, cocco2018inverse} -- a graphical model based approach -- to model the distribution of sequences subject to selection constraints. This inverse modelling approach consists in finding effective energetic parameters (fields and couplings in a Potts Hamiltonian) describing the empirical distribution of sequences. Applying DCA to the exactly solvable model of lattice proteins allows us to describe in great detail our dimeric system. We show that the inferred couplings are excellent predictors of the intra- and inter-protein structure; moreover, they indirectly keep track of the effect of selection due to the mentioned constraints.

In the second part we focus on the full evolutionary history, from the initial state with two non-interacting LPs in their native structures to the final state where a bounded dimer is formed. We look at the competitive dynamics between folding into the native structures vs. realizing the functional protein-protein interaction to characterize the evolutionary trade-offs due to the binding constraints. 

\section{The model}
\label{par:model}

\subsection{Native folds}
\label{par:model_native}

\begin{figure*}
\includegraphics[width=0.55\textwidth]{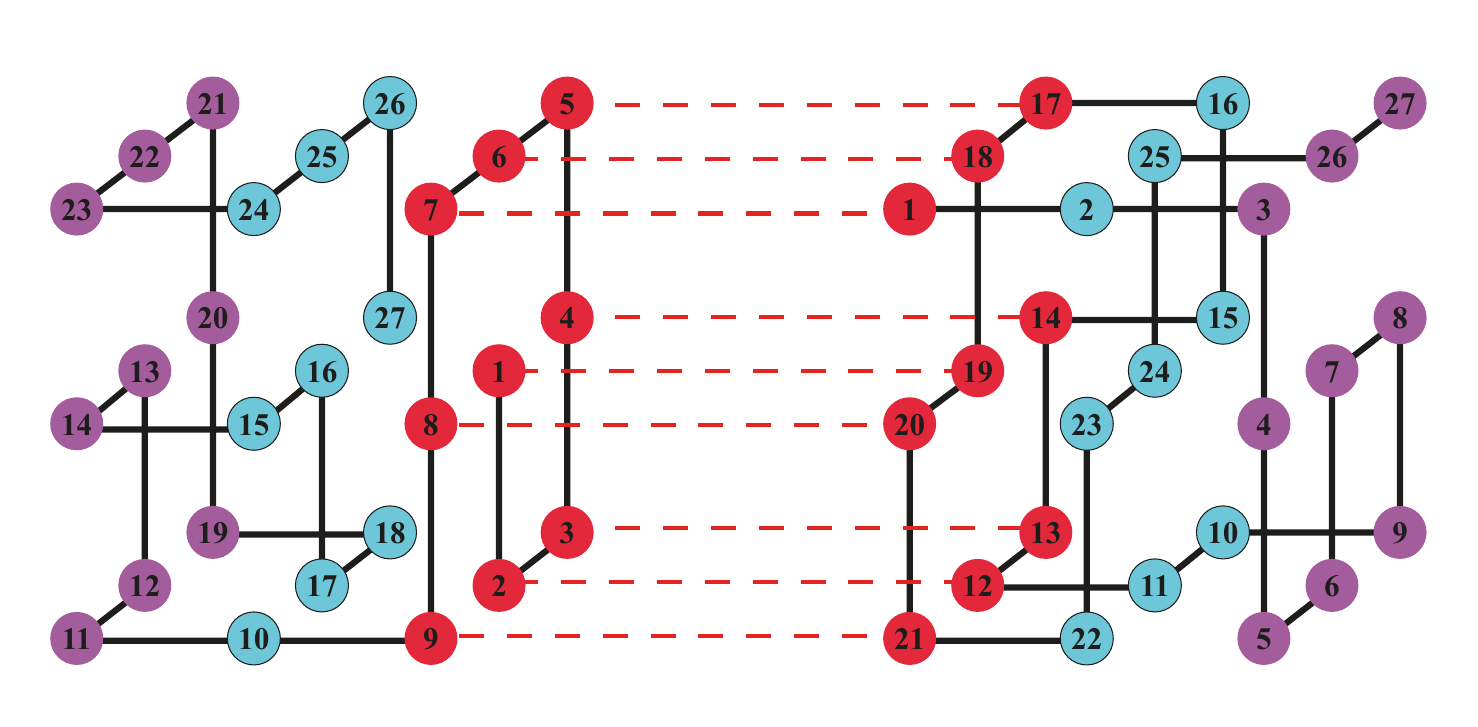}
\put(-235,-5){$\mathbf{S_C}$ \qquad \qquad  \qquad \qquad  \qquad \qquad \qquad $\mathbf{S_A}$}
\caption {\label{fig:dimer}Representation of the protein dimer studied in this work, composed of a structure $S_A$ (right) attached to a structure $S_C$ (left) through a specific face and orientation, which we refer to as functional binding mode (red dotted lines in \cref{fig:dimer}). For each structure the backbone of protein is highlighted with solid black lines. Dashed lines highlight binding contacts between the two structures.}
\end{figure*}

We focus on an exactly solvable model, namely lattice proteins (LPs), to study the formation of protein dimer, that is a macromolecular complex formed by two protein monomers. Each model protein consists of a chain of $L=27$ amino acids that occupy the sites of a $3 \times 3 \times 3$ cubic lattice \cite{shakhnovich1990enumeration,li1996emergence}. A valid conformation, hereafter called structure or fold, is a non-interacting chain that visits each site once, and there are $103,406$ of such possible structures (excluding global symmetries) \cite{shakhnovich1990enumeration}. Two examples of folds, called $S_A$ and $S_C$, are shown in \cref{fig:dimer}. For computational efficiency, we restrict ourselves to a representative subset of $\mathcal{N} = 10,000$ structures\cite{heo2011topology}.

Two amino acids are said to be in contact if they are nearest neighbours on the cubic lattice (but not on the backbone). The contact matrix $\vb{c}^{S}$ of structure $S$ is the $27 \times 27$ adjacency matrix such that 
\begin{equation}
\label{eq:cmap}
\begin{split} 
c_{ij}^S = \begin{cases}
1  \qquad i,j \quad \text{in contact}, \\
0 \qquad \text{otherwise}.
\end{cases}
\end{split}
\end{equation} 
This matrix fully defines the structure $S$. Given 
 a sequence $\vb{A} = (a_1, \dots ,a_{27})$ of amino acids folded into structure $S$, we can assign the energy 
\begin{equation} 
\label{eq:E_MJ}
\mathcal{E}(\vb{A} | S) = \sum_{i<j} c_{ij}^S E(a_i,a_j) \ ,
\end{equation}
where residues in contact interact via the Miyazawa-Jernigan potential $E$ \cite{miyazawa1985estimation}, an empirical, symmetric $20 \times 20$-dimensional matrix containing effective interaction energies for each couple of amino acids.
The probability that sequence $\vb{A}$ folds into structure $S$, hereafter called native probability, is given by the Gibbs-Boltzmann distribution at unit temperature, \ie{}
\begin{equation} 
\label{eq:pnat} 
P_{nat}(S |\vb{A} ) = \dfrac{e^{- \mathcal{E}(\vb{A} | S)}}{\sum_{S'=1}^{\mathcal{N}} e^{- \mathcal{E}(\vb{A} | S')} }.
\end{equation} 
Stable structures $S$ for the sequence $\vb{A}$ are the ones that  maximize the gap between their energy $\mathcal{E}(\vb{A} | S)$ and the one of competing structures $S'$.

\subsection{Protein-protein interaction}
\label{par:model_int}

We now move the attention to dimer lattice proteins (hereafter only referred to as dimer), \ie{} aggregate of two lattice proteins bounded together trough one of their faces, whose chain is made of a sequence of $L=54$ amino acids. To model protein-protein interaction we decide to look at only one among the $6 \times 6 \times 4 =144$ possible binding modes, so the two lattice proteins interact functionally via two specific faces with a specific orientation. Indeed, here one interface is considered \textit{functional} while the 143 remaining might exist but are not deemed as functional in this model. An example of dimer - which corresponds to the one we studied in this work - is shown in \cref{fig:dimer}.
The interaction between faces of structure $S_1$ and $S_2$ brings into the arena a new energy contribution in the form 
\begin{equation} 
\label{eq:E_int}
\mathcal{E}_{int} (\vb{A_1},\vb{A_2} | S_1 + S_2, k ) = \sum_{i \in S_1, j \in S_2} b_{ij}^{k} E(a_i,a_j),
\end{equation} 
where $\vb{A_1}, \vb{A_2}$ are the amino acid sequences folded in $S_1, S_2$ respectively, and $b_{ij}^{k}$ is the contact map of the $k$-th binding mode, namely 
\begin{equation}
\label{eq:cmap_binding}
\begin{split} 
b_{ij}^k = \begin{cases}
1  \qquad i \in S_1,j \in S_2 \quad \text{in interaction}, \\
0 \qquad \text{otherwise}.
\end{cases}
\end{split}
\end{equation} 
Hence, the interaction probability associated to the functional binding mode (hereafter always labelled with $k=1$) reads 
\begin{equation} 
\label{eq:pint} 
P_{int}(S_1 + S_2 | \vb{A_1}, \vb{A_2}) = \dfrac{e^ {- \mathcal{E}_{int} (\vb{A_1},\vb{A_2} | S_1 + S_2, 1 )} }{\sum_{m=1}^{144} e^ {- \mathcal{E}_{int} (\vb{A_1},\vb{A_2} | S_1 + S_2, m )} } \ .
\end{equation}

\subsection{Dimerization}
\label{par:model_dim}

We can now write the full probability for a dimer (up to a normalization constant):
\begin{equation}
\label{eq:pfull}
\begin{split}
P (\vb{A_1}, \vb{A_2})  &\propto P_{nat} (S_1 | \vb{A_1})^{ \hbnat}\times  P_{nat} ( S_2 | \vb{A_2} ) ^{\hbnat} \\
&\times P_{int} (S_1 + S_2 | \vb{A_1}, \vb{A_2})^{\hbint}
\end{split}
\end{equation}

The exponents $\hbnat$ and $\hbint$ acts as stressors that control the stringency of evolutionary selection to fold into native conformations and to bind functionally. In practice, we want $P_{nat}$ and $P_{int}$ to reach values very close to 1, {\em e.g.} 0.99 or 0.999, implying that  $\hbnat$ and $\hbint$ must be very large, of the order of $1000$. To avoid manipulating these large numbers we introduce the rescaled stressors
\begin{equation}
    \bnat=\frac{\hbnat}{1000}\quad , \qquad \bint=\frac{\hbint}{1000}\ ,
\end{equation}
where $\bnat$ and $\bint$ are of the order of 1.

The effective Hamiltonian of the two-sequence system therefore reads 
\begin{eqnarray}
\label{eq:effective_hamiltonian}
        \mathcal{H}(\vb{A_1}, \vb{A_2})  &= &- 1000\, \bnat\, \log P_{nat}( S_1 | \vb{A_1} )  \\
        &-& 1000\,\bnat \,\log P_{nat} (S_2 | \vb{A_2}) \nonumber \\
        &-& 1000\,\bint\, \log P_{int} (S_1 + S_2 | \vb{A_1}, \vb{A_2} )
         \ . \nonumber
\end{eqnarray}

\section{Sampling dimer space}
\label{par:sampling}

\subsection{Monte Carlo Metropolis sampling}
\label{par:MC_sampling}

We generate a multiple sequence alignment (MSA) for the dimer formed by two lattice proteins folded in two specific structure, labelled with $S_A$ and $S_C$ as in \cite{jacquin2016benchmarking}, through Monte Carlo (MC) simulations via the Metropolis rule. For each dimer in the MSA, the simulation starts with two sequences folded in structure $S_A(=S_1)$ and $S_C(=S_2)$ randomly taken from the two MSAs built in \cite{jacquin2016benchmarking}. At each time step we perform the following routine to update the sequences:
\begin{enumerate}
    \item \textbf{Mutation:} a move consisting of mutating at random two amino acids (one on each protein) is attempted - say, $a_{1i} \to a'_{1i}$ and $a_{2i} \to a'_{2i}$; let $\vb{A_1 ', A_2 '}$ denote the mutated dimer, while $\vb{A_1 ,A_2}$ refers to the old dimer. 
    \item \textbf{Selection:} If $P (\vb{A_1 ', A_2 '}) > P(\vb{A_1 ,A_2})$ the move is accepted, otherwise it is accepted with probability $ P (\vb{A_1 ', A_2 '}) / P (\vb{A_1 ,A_2})$.
\end{enumerate}

 For each couple of value $(\bnat, \bint)$ we construct a MSA of $N = 23,000$ dimer sequences, separated by  $N_{it} = 5,000$ MC steps to avoid correlations. The time taken by the dimer to relax to equilibrium is independent from the two  initial  sequences and slightly varies on the values $(\bnat, \bint)$, but remains smaller than $N_{it}$ steps. Moreover, in order to study the evolutionary dynamics of dimer we also built MSA at intermediate steps, {\em i.e.} before reaching thermalization. In practice, during MC simulations, we collected MSA of evolving dimer at steps  $0;50;200;500;1,000;3,000;5,000$ - with step $= 0$ corresponding to the starting situation, while step $=N_{it}$ corresponds to the final dimer once thermalization has occurred. 

\subsection{Evolutionary model sampling}
\label{par:evolutmodel_sampling}

An alternative way to construct a MSA dimer is to simulate the evolutionary dynamics of a population of individuals (sequences), hereafter referred to as $\vb{X}$. Having an additional MSA generated with a completely different approach allows us to perform the analysis on both of them independently and check whether we obtain the same results. 

The evolutionary model starts with a population of size $N$, where each individual has a genome made by the two LP chains. The initial sequences  have high folding probabilities but do not generally interact, {\em i.e.} they  were generated through the previous MC dynamics with $\bnat=1, \bint=0$. For the sake of simplicity we then randomly pick up one of the sequences $\vb{A_2}$  for $S_C$ and freeze it. The evolutionary process will focus on sequences $\vb{A_1}$ for $S_A$ only. 

At each generation of the evolutionary dynamics, two steps are carried out to make evolve the parent population $\vb{X}$ into a new population, $\vb{X^*}$:
\begin{enumerate}
    \item \textbf{Mutation:} for each individual and for each site of the sequence  we draw a binomial random variable $m_i=0,1$ with mean  $\mu$. If $m_i=1$ the amino acid on site $i$ is replaced with a new one, drawn from a background distribution of frequencies of amino acids in the MSA associated to structure $S_A$. We end up with a new population $\vb{X'}$.

    \item \textbf{Selection:} to each individual $X'_i=(\vb{A_1},A_2)$ in the mutated population $\vb{X'}$ we associate a fitness value $P(X'_i)$ given by \cref{eq:pfull}. Then we draw a multinomial random variable $\kappa = 1,\dots, N$ according to the  fitnesses of all individuals in the population, \ie{} the probability of drawing $\kappa$ is
    \begin{equation}
        \mathcal{P}_{\kappa} = \dfrac{P(X'_{\kappa})}{ \sum_i P(X'_i)}.
    \end{equation}
    We thus generate an off-spring identical to $\kappa$-th mutated individual in $\vb{X'}$. This random extraction process is repeated $N$ times (with replacement), and we end up with the population of off-springs $\vb{X^*}$, where the numbers of copies of individuals in  $\vb{X'}$ are, on average, proportional to their fitnesses.
\end{enumerate}

\noindent These two steps are repeated $T$ times, and results are averaged over multiple sample populations.

\section{Equilibrium properties for various stressor strengths}
\label{sec:Equilibrium_prop}

In this section we investigate how the value of the stressor $\bint$ affects the sequence distributions after evolution has reached an equilibrium state. The dynamical transient preceding this equilibrium regime will be studied in the next section.

The study of equilibrium properties of the dimer can be carried out from sequence data generated either with MC sampling or population dynamics, with similar results from a qualitative point of view. Hereafter, we use MC generated data to characterize the statistics of residues in the dimer and its connection with structural properties (inter- and intra-protein contacts). We then focus on sequences produced by the population dynamics model to understand the interplay between the population size, the number of mutations per individual, and the response to the stressor.

Unless otherwise said, we perform simulations sharing the same parameter $\bnat=1$  for the two models. We also set for population dynamic model $\mu = 1/L$, so that, on average, we have one mutation per individual as in the MC evolution. The population dynamics model is characterized by an additional parameter to tune, namely, the population size.

\subsection{Foldability--dimerization trade-offs, and their effects on sequence statistics}
\label{par:dimer_characterization}

\begin{figure}
\centering
    \includegraphics[width=\columnwidth]{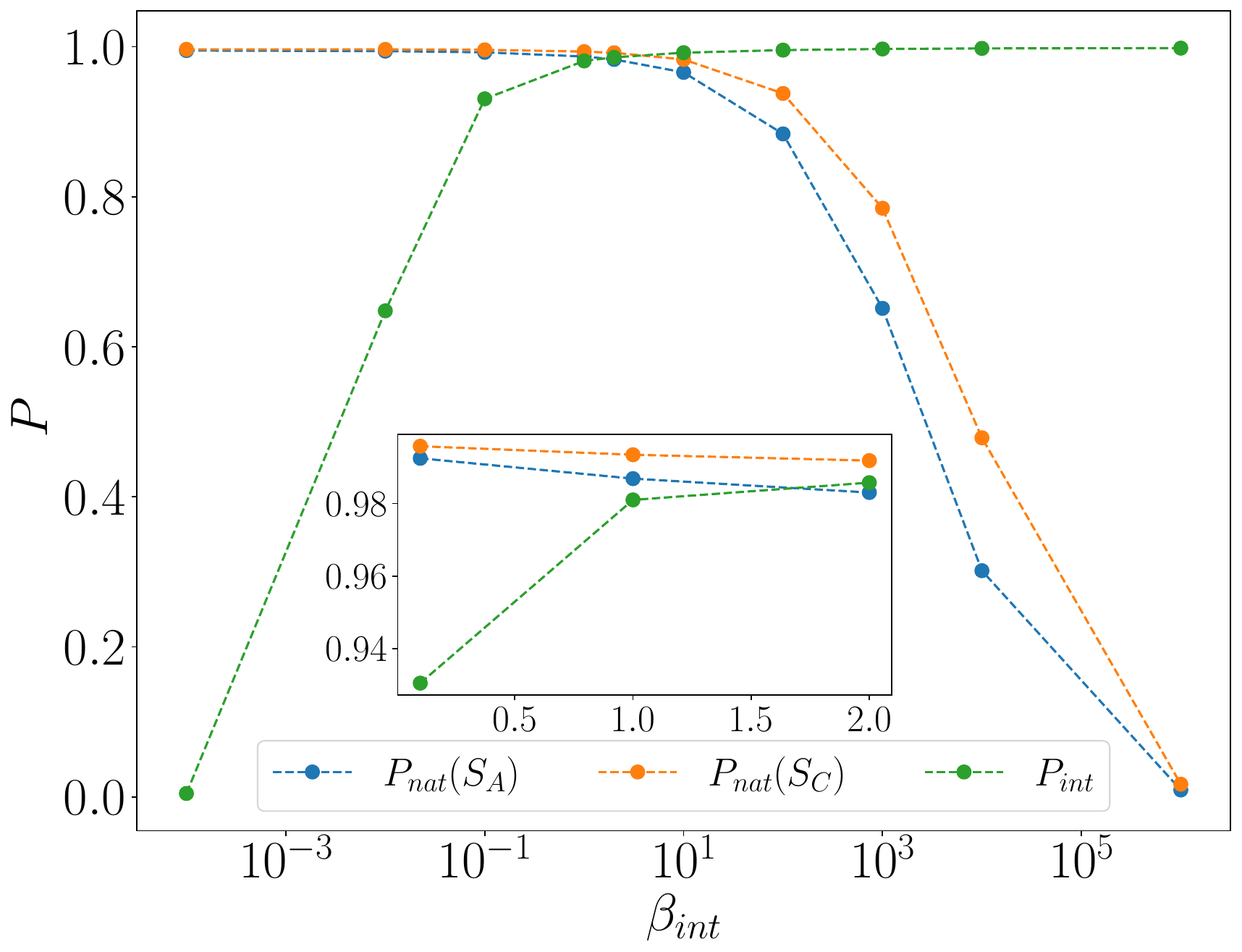}
    \caption{\label{fig:pnatpint_vs_gamma}Values of $P_{int}$, $P_{nat} (S_A)$, $P_{nat}(S_C)$ as functions of $\bint$ at equilibrium. Inset: zoom in the region $\bint \sim 1$, where all probabilities have high values. Averages are carried out over $10^4$  realizations.}
\end{figure}

We start by tuning $\bint$ to see how changing the selection pressure applied along the $P_{int}$ direction affects the capability of the individual proteins to reach their native folds. The behaviours of $P_{nat}(S_A|\vb{A_1})$, $P_{nat}(S_C|\vb{A_2})$, and $P_{int}(S_A+S_C|\vb{A_1,A_2})$ are shown in \cref{fig:pnatpint_vs_gamma}. As a general trend, $P_{int}$ is an increasing function of $\bint$ as expected from the explicit dependence of $P$ on $\bint$ in \cref{eq:pfull}. We also observe that the two $P_{nat}$ decrease with $\bint$, an effect of the evolutionary trade-offs between two competing fitness components. As $\bint$ grows to large values, the probabilities that the two proteins adopt their native folds become small: in this regime the selective pressure favoring protein-protein binding is too strong to cope with the folding constraints. For intermediate values of $\bint$ we are able to obtain both good foldings and high interaction between the two structures.  

The presence of trade-offs observed at the phenotypic level in \cref{fig:pnatpint_vs_gamma} can be studied at the sequence level. To obtain a fine characterization of the sequence statistics resulting from the evolutionary constraints, we infer a pairwise Potts model with $ q= 20$-state variables (corresponding to the $20$ possible amino acids), and $N= 54$ variables (corresponding to the full dimer length) \cite{wu1982potts}. The energy of this effective model is the sum of two contributions, featuring local and interacting terms: \begin{equation}
\label{eq:e_potts}
    \mathcal{H}^{\text{Potts}}(\vb{A_1,A_2}) = - \sum_{i <j} J_{ij}(u,v) \delta_{a_i,u} \delta_{a_j,v} - \sum_i h_i (v) \delta_{a_i,v},
\end{equation}
where the indices $i,j$ run along all the dimer sequence, {\em i.e.} $i, j = 1, \dots, L$.
Here, $J_{ij} (u,v)$ represents the coupling between amino acid $u$ at position $i$ and amino acid $v$ at position $j$ along the sequence; given two positions $i,j$ in contact in the 3D dimer structure, they reproduce the interacting MJ potential $E(u,v)$. The parameters $h_i (u)$ are local fields acting on position $i$ that depends on specific amino acid $u$. To infer the values of these parameters  we follow a Boltzmann Machine (BM) learning procedure \cite{tubiana2019learning}, consisting in maximizing the average log-likelihood $\expval{\log \mathcal{P}}_{\text{data}}$ of the model over the sequence data. In practice, we assume the data to be Gibbs-distributed as
\begin{equation}
\mathcal{P}^{\text{Potts}} = \dfrac{e^{-\mathcal{H}^{\text{Potts}}}}{Z},
\end{equation} 
where $\mathcal{H}^{\text{Potts}}$ is given by \cref{eq:e_potts}. We compute $\expval{\log \mathcal{P}}_{\text{data}}$ over the MSA dataset for an initial guess of the parameters; we then proceed to its maximization by numerically ascending the gradient $\nabla \expval{\log \mathcal{P}}_{\text{data}} $, until the log-likelihood is maximized 
(see \cref{par:methods} for further details on BM learning). The meaning of the inferred inter- and intra-protein couplings is studied below in \cref{sec:structural_significance} and \cref{sec:dimer_characterization}. \\
A stringent test of the accuracy of the  inferred Potts model is its ability to generate new dimer sequences that have both the right target native structures and high, specific face-face interaction. In practice, we generate sequences with MC from the inferred distribution
\begin{equation}
    {\cal P}^\text{Potts}(\vb{A_1,A_2}) \propto e^{-  \mathcal{H}^{\text{Potts}}(\vb{A_1,A_2})/T}\ ,
\end{equation}
where $T$ is a fictitious sampling temperature used to control the broadness of the sampled region, hence the diversity of the generated MSA. By choosing $T$ lower than unity, which is the implicit value of the inference temperature, we are able to sample sequences with low energies. For further details about the generation procedure we refer to \cref{par:methods}. 

\begin{figure}
    \centering
    \includegraphics[width=\columnwidth]{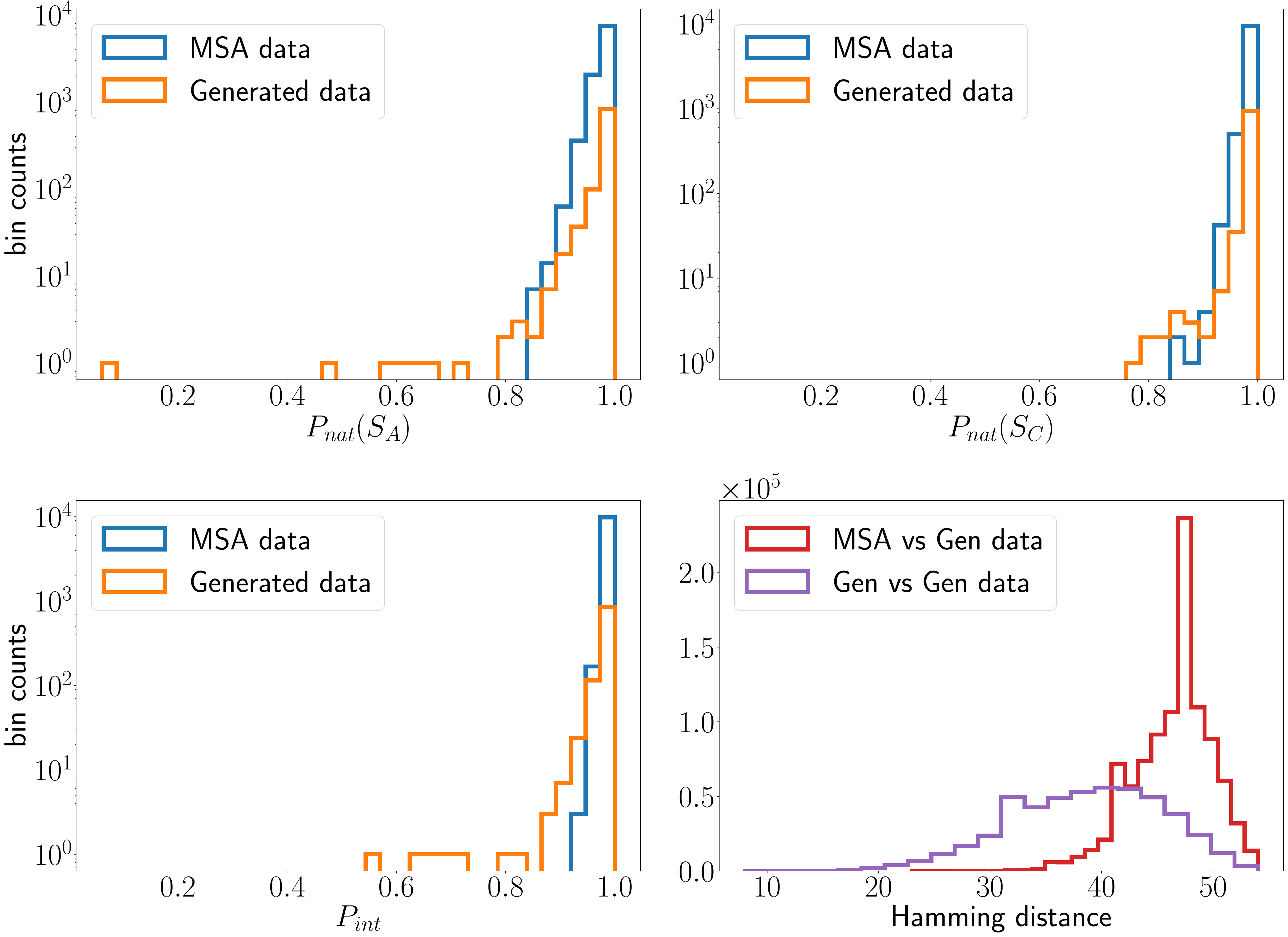}\caption{\label{fig:distribution_gen_train}Distribution of $P_{nat}(S_A)$, $P_{nat}(S_C)$, $P_{int}$ for the MC MSA and generated MSA (blue and orange, respectively). Bottom rigth: distribution of Hamming distance between the MC MSA and generated MSA and inside the generated MSA (red and purple, respectively). We sample new sequences with $T=0.5$.}
\end{figure}

We show in \cref{fig:distribution_gen_train} the histograms of the ground-truth probabilities --\cref{eq:pnat} and \cref{eq:pint}-- which show these sequences are good dimers. We see that these probabilities are quite high, proving that most of the necessary information needed to model a good dimer can be captured by pairwise interactions and local biases, in agreement with some recent works \cite{marmier2019phylogenetic, gandarilla2020statistical}.  In addition, the generated sequences are far from the old ones and have high diversity between each other, as it can be seen in \cref{fig:distribution_gen_train} (bottom right panel) where we plot the distribution of the Hamming distance for each pair of sequence; hence we are sampling a different subset of the dimer space. 

\subsection{Structural significance of the statistical couplings}
\label{sec:structural_significance}

Following the standard direct-coupling approach \cite{cocco2018inverse}, we rank the inferred couplings in descending order of their Frobenius scores ($L_2$ norms), 
\begin{equation}
    F_{ij} = \sqrt{\sum_{a_i, a_j} J_{ij} (a_i,a_j) ^2} ,
\end{equation}
and use this ranking as a predictor of the contacts in the dimer. Informally speaking, we expect the strongest inferred couplings to be those corresponding to real contacts among amino acids in the dimer (see \cref{par:methods}). We compute the positive predictive value (PPV)  at rank $k$ as the fraction of the top-$k$ scores whose pairs of positions along the sequence are effectively in contact in the 3D dimer. 

\begin{figure*}
    \includegraphics[width=\textwidth]{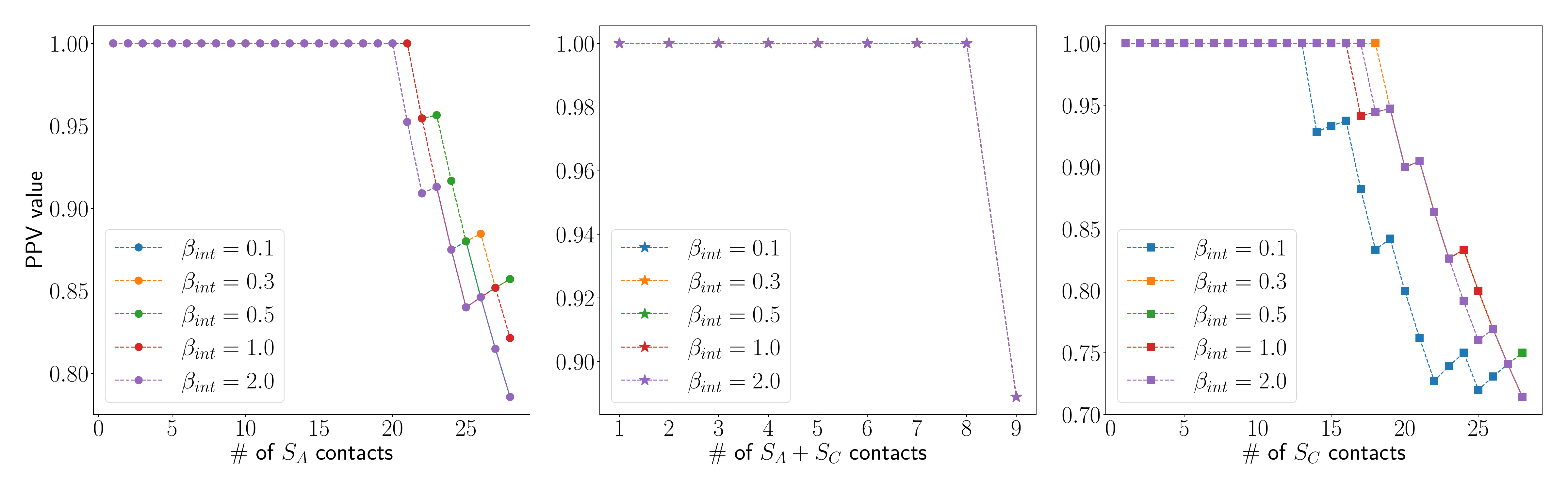}
    \caption{PPV for structure $S_A$ (left panel, circles) and structure $S_C$ (right panel, squares) at different values of $\bint$. Middle panel: PPV for contacts belonging to the functional binding mode (stars). Data points at different $\bint$ overlap, as we are able to predict all contacts but the central one. The MSA is made of $23,000$ dimer sequences evolved for $5,000$ MC steps. }
    \label{fig:ppv}
\end{figure*}

We show in \cref{fig:ppv} the PPV used for contacts prediction in structure $S_A$ and $S_C$ (left and right panels, respectively), for different values of the interaction strength $\bint$. Between the two structures, we can see that - whatever the value of $\bint$ is - the PPV performs better on predicting contacts in structure $S_A$ rather than in $S_C$. Indeed, for $S_A$ we always predict the first $19$ contacts out of the total $28$, while for $S_C$ we miss more contacts.
The missed contacts always include the central site and the central site of the binding mode, see below.
Both structures $S_A$ and $S_C$ can host an extremely large number of sequences: the higher this number, the more designable the structure is said to be. However, it appears that having high designability is harmful for inferring couplings: between $S_A$ and $S_C$, the former can host less sequences, which means it is more specific and allows for better inference and contacts prediction. As for the different interaction strengths, we cannot identify a particular trend for increasing values of $\bint$ both for structure $S_A$ and structure $S_C$. 

We also compute the PPV for the binding mode for several values of $\bint$ in order to assess the quality of contact prediction between the two structures (see \cref{fig:ppv}, middle panel). Regardless of the interaction strength, we only miss one contact out of the nine present on the interacting layer. Interestingly, the missed contact is always the same for all $\bint$, and corresponds to the central contact between sites $1-19$ in \cref{fig:dimer}. In fact, contacts involving central sites are generally the ones predicted worse, as central sites have more neighbours and  are often present in several competing structures \cite{jacquin2016benchmarking}. 

\subsection{Characterization of dimer interactions}
\label{sec:dimer_characterization}
\begin{figure*}
    \centering
    \includegraphics[width=\columnwidth]{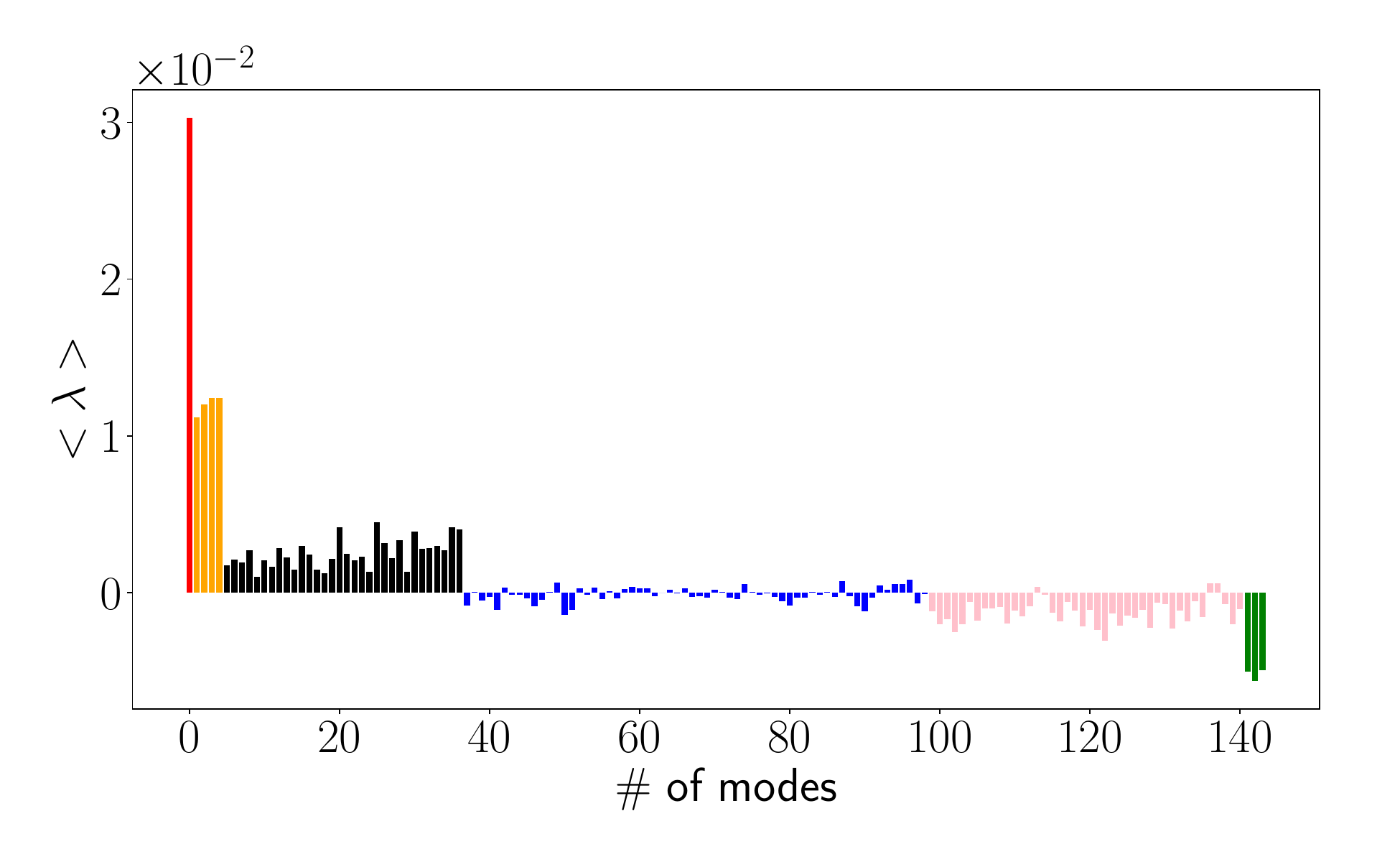} 
    \includegraphics[width=\columnwidth]{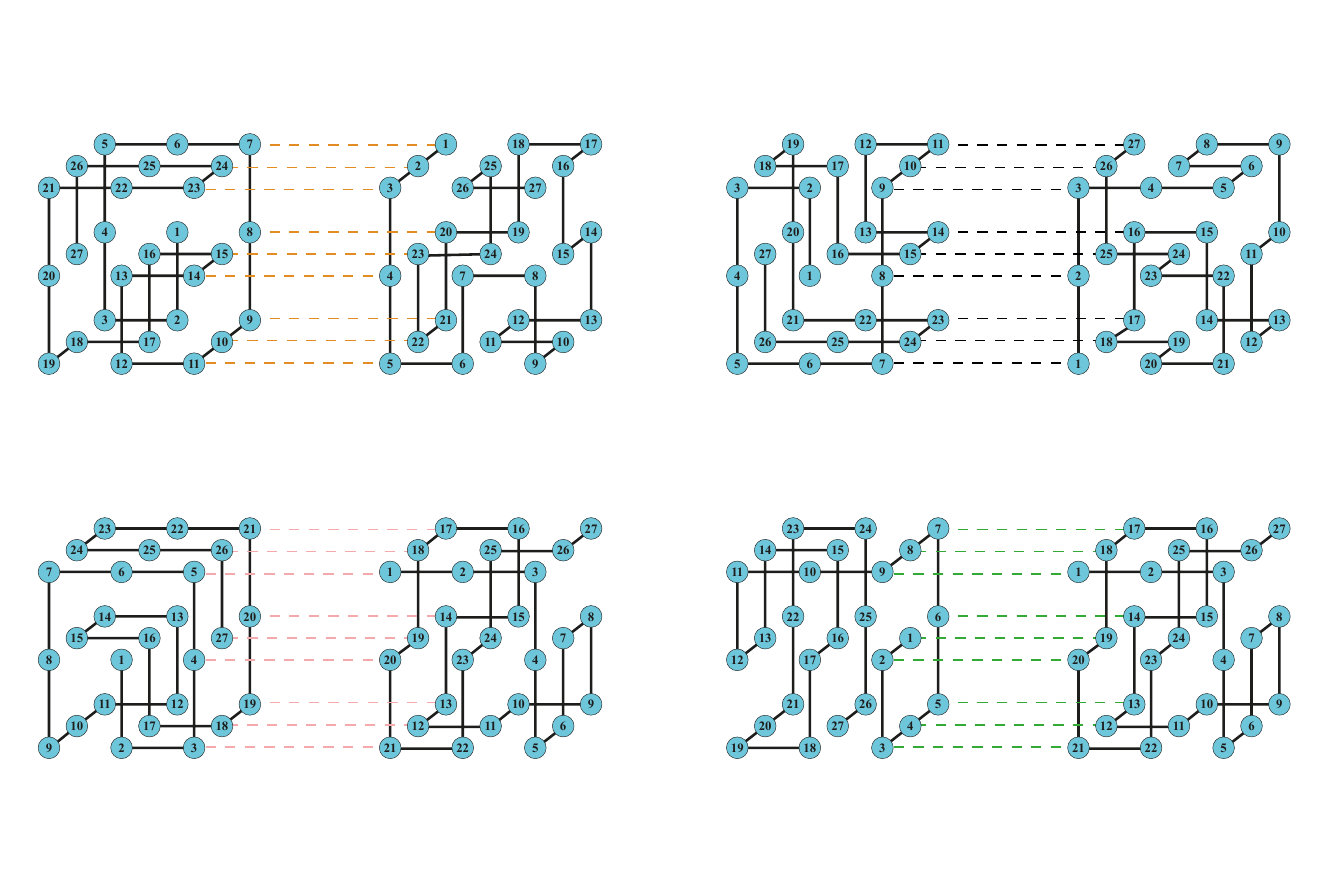}
    \put(-250,150){a)}
    \put(-120,150){b)}
    \put(-250,70){c)}
    \put(-120,70){d)}
    \caption{Left: Scores $\lambda_{ij}$ averaged over all the nine binding contacts for each mode, computed at equilibrium according to \cref{eq:projections}. Except for the scores associated to binding modes described in the main text (see \cref{sec:dimer_characterization} for orange, black, pink and green bars and \cref{fig:dimer} for the red bar related to functional binding), the remaining binding modes - the $62$ blue ones - exhibit a rather flat landscape. They are associated to binding modes that do not involve sites belonging to the interacting faces of the functional binding mode (\eg{} binding between the two back faces), hence they are just slightly favored or disfavored depending on the specific case. \\
    Right: Schematic of representative binding modes belonging to each of the subsets described in the main text: a) orange mode, b) black mode, c) pink mode, d) green mode.}
    \label{fig:projections}
\end{figure*}

Additionally, we characterize the binding modes between the two structures using the quantity $\lambda_{ij}$ - which is another score computed again from the inferred couplings $J_{ij}$ that we used to assess the quality of the functional binding mode against the remaining $143$ modes. Indeed, in \cite{jacquin2016benchmarking} the authors observe a linear dependency between the couplings and the MJ energy matrix, with a slope given by $\lambda_{ij}$ for each pair $(i,j)$. Such slope can be computed as
\begin{equation}
\label{eq:projections}
    \lambda_{ij} = - \dfrac{\sum_{a,b} J_{ij} (a,b) E(a,b)}{\sum_{a,b} E(a,b)^2},
\end{equation}
where the sum runs over all the possible amino acids. The quantities $\lambda_{ij}$ can then be seen as a measure of the coevolutionary constraints imposed by the design of the two structures.\\
The projection scores $\expval{\lambda}$, averaged over the nine binding contacts for each binding mode, are shown in \cref{fig:projections} (left panel). Among them, we have identified some relevant subsets depending on the number of maintained/mismatched contacts involving the interacting faces wrt the functional binding mode contacts (see \cref{fig:projections}, right panel for a visual representation with the associated color code):
\begin{itemize}
    \item single red peak corresponds to the functional binding and it is the highest one;
    \item four orange peaks corresponding to binding modes where three functional contacts are still maintained. An example of such configuration is in \cref{fig:projections}, right panel a);
    \item black subset corresponds to binding modes where one out of nine binding contacts is still present. An example of such configuration is in \cref{fig:projections}, right panel b);
    \item pink subset contains all binding modes where there is at least one mismatched contact. An example of such configuration is in \cref{fig:projections}, right panel c);
    \item three green binding modes that have eight out of nine mismatched contacts. An example of such configuration is in \cref{fig:projections}, right panel d).
\end{itemize}

In the four orange configurations, the scores $\lambda_{ij}$ associated to the three binding contacts of the functional mode are very high and thus are responsible for a large value of $\expval{\lambda}$. The same holds for black configurations, where this time only one $\lambda_{ij}$ is high. Surprisingly the orange peaks are almost $1/3$ of the red one and the black ones are on average almost $1/9$ of the red one.\\
Conversely, the pink and green configurations have, respectively, one and eight negative scores thus resulting in binding modes that we strongly avoid.

\section{Transient responses to stressor }
\label{par:evolutionary_tradeoff}

Up to now we have only discussed the stable state of the dimer sequence distribution. We now consider the effects of a rapid change of the stressor value $\bint$. 

\subsection{Dynamical recovery of structural fitness}

\subsubsection{Response to a step-like change}

In order to achieve a clear view of what happens during MC evolution, we compute the native probability $P_{nat}$ and the interaction probability $P_{int}$ at each time step for the whole length of the MC simulation and for $N_{seq}=1000$ dimer sequences. We then average $P_{nat} (t)$ and $P_{int} (t)$ over these $N_{seq}$ dimer sequences.

\begin{figure}
\centering
    \includegraphics[width=\columnwidth]{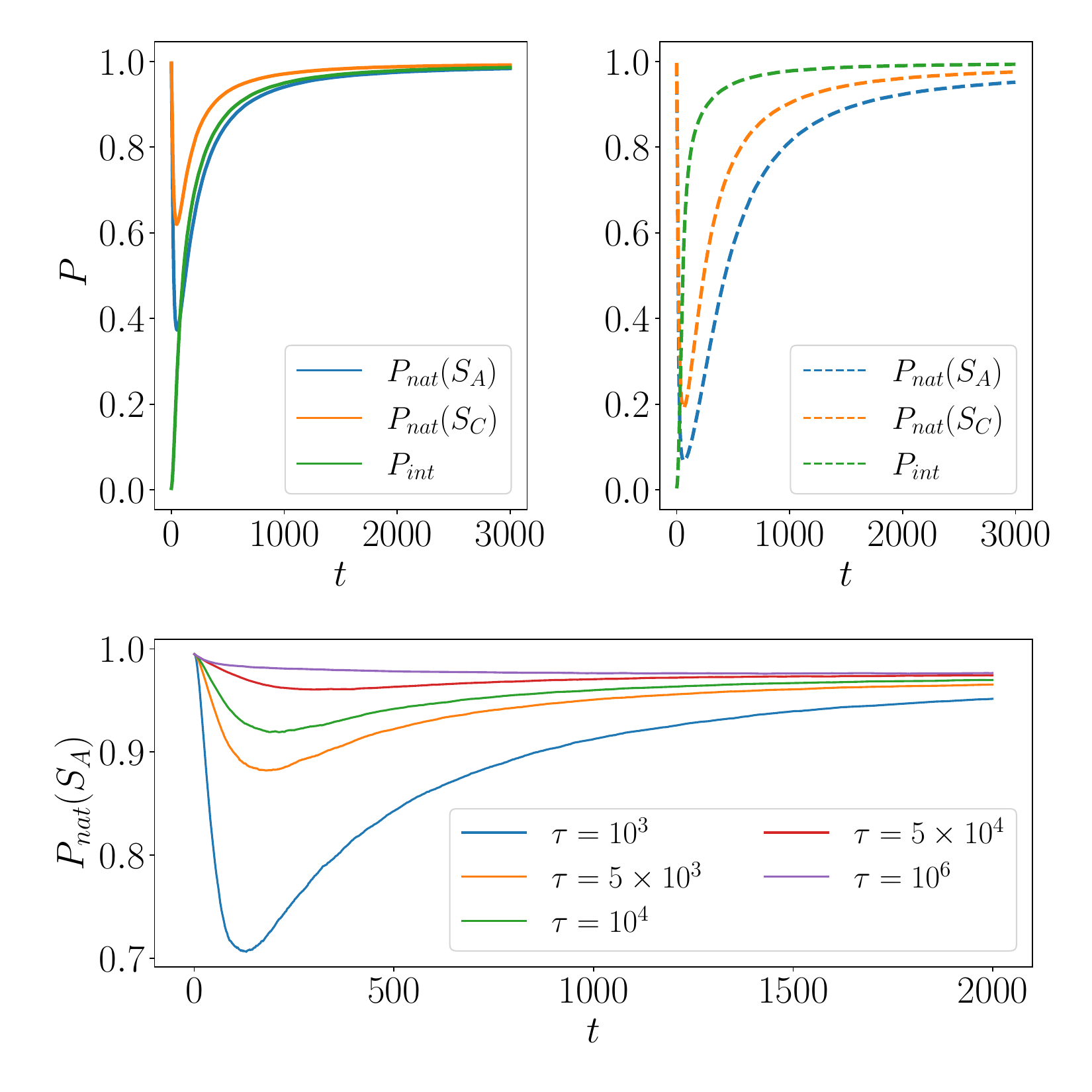} 
    \caption{\label{fig:evolution}Top: Evolutionary trajectories of $P_{nat}$ for $S_A$ (blue) and $S_C$ (orange) and of $P_{int}$ (green) for weak (left, solid line with $\bint = 0.5$) and strong (right, dotted line with $\bint = 10.0$) protein-protein interactions. Curves are averaged over $3,000$ dimer sequences. Bottom: Evolutionary trajectory $P_{nat} (S_A)$ for increasing values of $\tau$. The smaller $\tau$, \ie{} the sharper is the $\bint(t)$ dependence, the more we experience the out of equilibrium effect. Let us note that here we are plotting the short-time evolution, since we are only interested to see how the sharpness of the stressor $\bint$ affects the evolution. Here exceptionally $\bnat = 0.2$ and $\bint(0)=0.0$, $\bint^F = 2.0$.}
\end{figure}

\cref{fig:evolution} (top) shows the evolution of $P_{nat} (t)$ and $P_{int} (t)$ in time for two values of $\bint$. For large $\bint$  we observe a huge drop in the folding  probabilities $P_{nat}$ associated to both structures $S_A$ and $S_C$, while $P_{int}$ monotonically increases to reach a $\bint$-dependent plateau value. This dynamical evolution is a direct consequence of the evolutionary trade-offs between  $P_{nat}$ and $P_{int}$, as the amino acids on the interacting faces enter both in \cref{eq:pnat} and in \cref{eq:pint}. The evolutionary trajectories in \cref{fig:evolution} show that, in order to increase the interaction probability associated to the binding mode, the two sequences are forced to go through sub-optimal - even very bad - states from a structural point of view  ($P_{nat}$ values down to $0.2$). After this strong drop and the amino acids controlling the dimer interaction have been optimized enough both $P_{nat}$ start increasing again but reach values lower than the ones at the beginning of the evolutionary trajectory, {\em i.e.} in the absence of interaction. The constraint arising from the binding mode does not allow them to optimally maximize their single structure folding; in addition, as expected, the larger $\bint$, the bigger the drop in $P_{nat}$ is for both structures, and the lower the final value of $P_{nat}$ \cite{peleg2014evolution}. 

Interestingly, for any $\bint$ value, the drop in $P_{nat}$ for structure $S_A$ is always larger than the one for $S_C$. Even if both structures are undergoing roughly the same number of mutations, structure $S_C$ remains more stable compared to $S_A$, a fact related to its larger designability \cite{jacquin2016benchmarking}. We will study this point in more details in section \ref{par:designability}.

\subsubsection{Case of smooth increases of the stressor strength}

The drop in $P_{nat}$ seen in \cref{fig:evolution} is an out-of-equilibrium effect, resulting from the abrupt change of selection pressure from 0 to $\bint$. To better study this effect, we consider a smooth, time-dependent stressor during the MC evolution 
\begin{equation}
\label{eq:gamma protocol}
    \bint (t) = \bint^F \; \tanh\left( \frac{t}{\tau}\right),
\end{equation} 
where $\tau$ sets the time scale of the stressor ($\tau \to 0$ gives back the step-like function studied so far). In \cref{fig:evolution} (bottom panel) we plot the transient dynamics of $P_{nat}$ for several values of $\tau$ with $\bint(t=0)=0$ and $\bint(t \to \infty) = \bint^F = 2.0$. While for long enough simulations, \ie{} with $t \gg \tau$, all curves reach the same plateau, the drop in $P_{nat}$ decays with the time scale $\tau$ in \cref{eq:gamma protocol}. For large enough $\tau$, \ie{} when the evolution can be considered adiabatic, $P_{nat}$ decreases monotonically over time.

\subsubsection{Evolution of a non-clonal population}
\label{par:non_clonal_pop}
While we have so far studied the evolution of a single sequence, we now consider the case of population of constant size $N$. Our aim is to characterize  how the selection pressure $\bint$ affects the substitution rate, the diversity of population, and also how the effects of the stressor relates to the population size. 

\begin{figure}
    \centering
    {\includegraphics[width=\columnwidth]{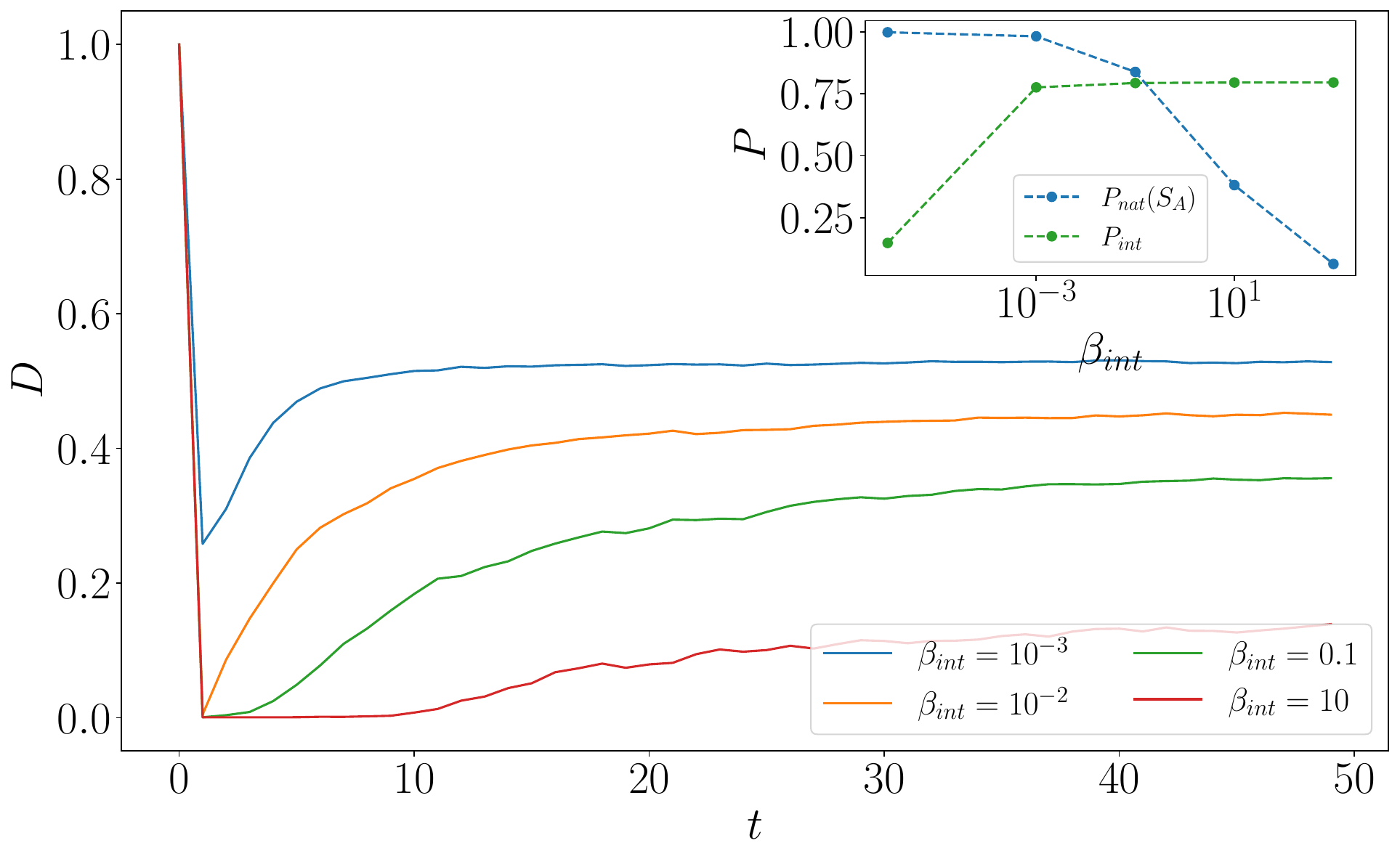}}
    \caption{Time course of the diversity $D(t)$ for different selection pressures $\bint$.
    Inset: fitnesses $P_{nat}(S_A)$, $P_{int}$ at equilibrium after population dynamic evolution for several stressor values $\bint$. We recall that $P_{int}$ can not increase further as we are keeping $S_C$ fixed compared to MC evolution. The parameters used are $\bnat=10^{-3}$, $N=1500$ and $\mu=1/L$, with $\bint$ kept fixed to its value during the evolution. Note that the values of $\bint$ are much smaller than in MC dynamics, due to the large population size, see \cref{sec:conclusion}.}

    \label{fig:evo_diversity}
\end{figure}

We first focus on the diversity $D(t)$, as the fraction of diverse individuals (unique sequences) present in the population at time $t$. The time behaviour of $D(t)$ is shown for various values of $\bint$ (at fixed size and mutation rates) in \cref{fig:evo_diversity}. Increasing $\bint$ makes the population less diverse, as fewer sequences satisfy the selection constraints and give rise to off-springs. The loss of diversity is maximal at the drop in $P_{nat}$. The inset of \cref{fig:evo_diversity} shows the stationary values of $P_{nat}$ and  $P_{int}$; we recall that the latter can not reach as high a value as with MC evolution since the second sequence in the dimer (associated to protein $S_C$ is not allowed to evolve).

\begin{figure}
    \centering
    {\includegraphics[width=\columnwidth]{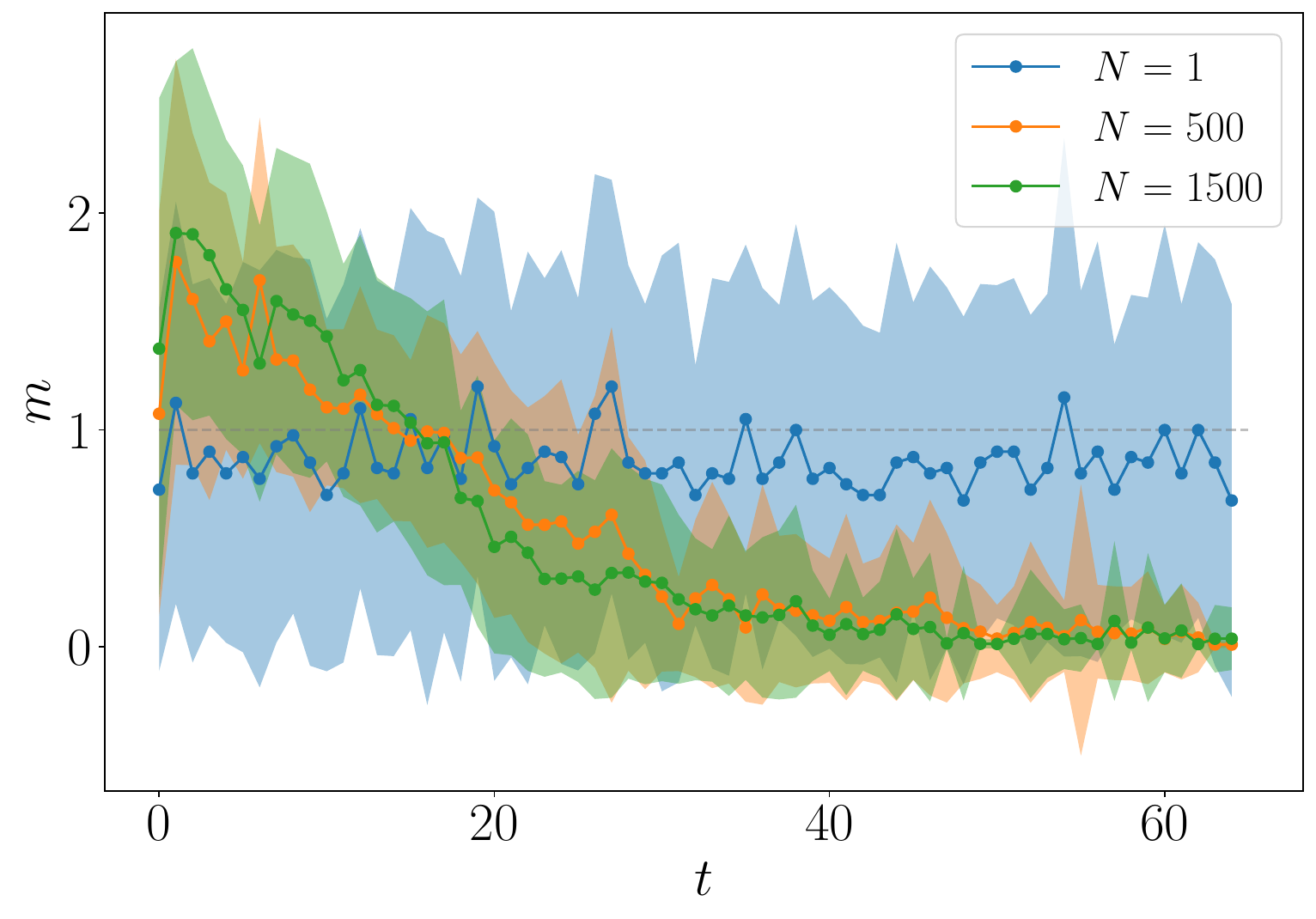}}
    \caption{Substitution rate $m(t)$ for different population sizes $N$.   The parameters used are $\bint = 10$ and $\mu=1/L$.}
    \label{fig:evo_substitution}
\end{figure}

We then study the substitution rate, $m(t)$, defined as the average Hamming distances between the  sequences at step $t+1$ and their parents at time $t$; without selection this rate would be on average equal to the number of mutations proposed per individual, {\em i.e.} $\mu L$. The substitution rate per individual is plotted as a function of the evolutionary time in \cref{fig:evo_substitution}. 
We observe, for sizes $N>1$, a peak in $m$ at the beginning of dynamics, decaying to a plateau. Conversely, for $N=1$, the substitution rate fluctuates around the average value $\expval{m}= \mu L$. The  initial peak in $m(t)$ is therefore mostly due to selection, rather than  to mutations (we recall that, in MC, $N=1$ and mutations are always proposed, an analogous quantity to $m$ being the acceptance rate).

\begin{figure*}
    \centering
    {\includegraphics[width=\textwidth]{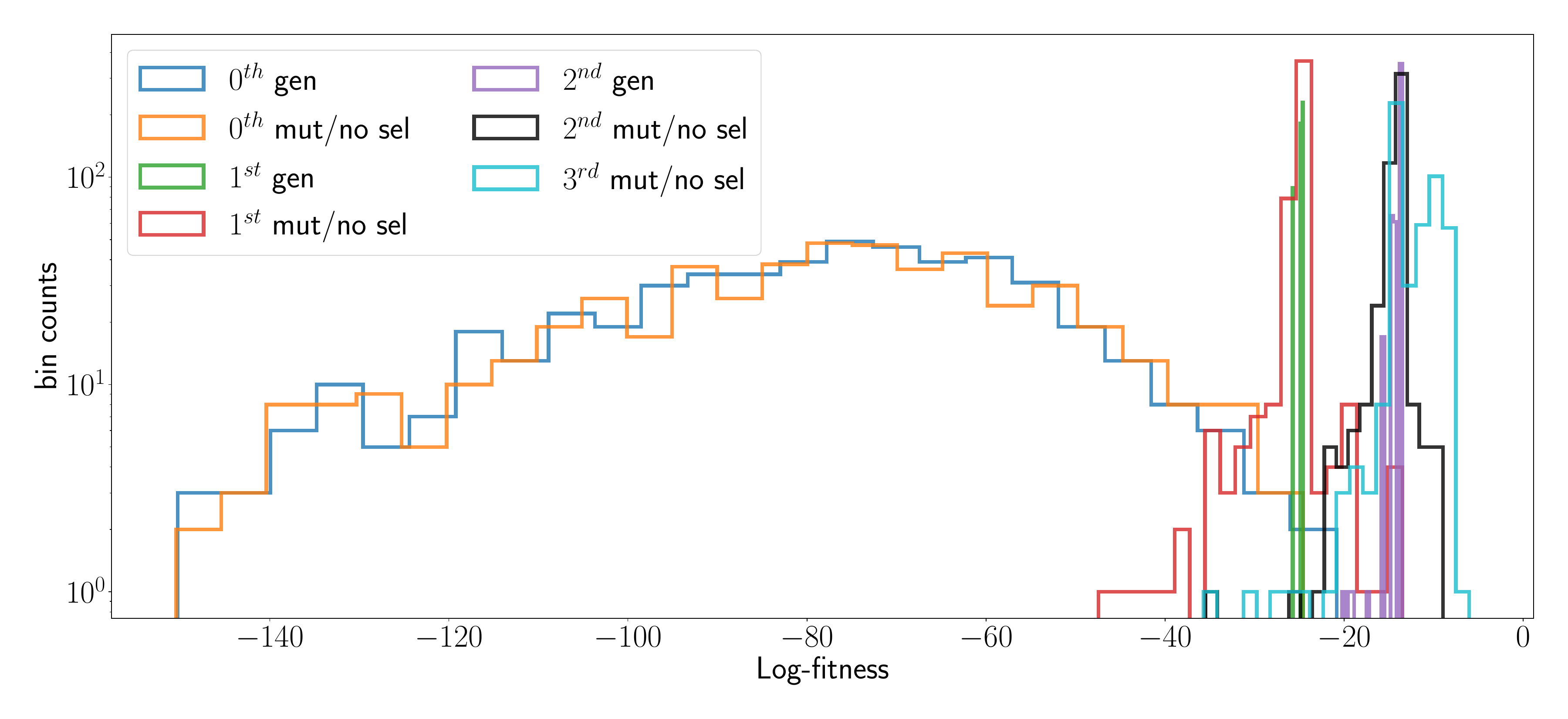}}
    \caption{Distribution of log-fitnesses during the first three steps of evolution. The parameter values are $\bnat=10^{-3}$, $\bint = 10^{-2}$, $\mu = 1/L$ and $N=1000$.}
    \label{fig:fitness_profile}
\end{figure*}

The time behaviours of the diversity and of the substitution rate can be qualitatively understood in a simplified scenario, in which the maximum of $m$ is achieved in only one evolutionary time point (cf. \cref{fig:fitness_profile} that supports our argument):
\begin{itemize}
    \item At $t=0$, all sequences are distinct, hence $D=1$, and the fitness distribution is very broad. This broad profile is maintained after random mutations are introduced. 
    \item Hence, at $t=1$, selection will only keep the few strains that are fitter in the population, resulting in a poorly diversity $D$. The fitness distribution is now strongly concentrated. Under mutations,  the distribution widens (and shifts to the left, since on average, there are more deleterious than beneficial mutations). \item At $t=2$ selection step picks just such mutated sequences that increased fitness, and as a result the substitution rate $m_{1 \to 2}$ is high. The diversity $D(2) > D(1)$ and the fitness profile is less peaked than before. 
    \item The mutation step does not impact the distribution of fitness,  and we expect $m(3) < m(2)$. At later times, the substitution rate will decreases and the diversity increases until both reach their plateau values. Fluctuations in $m(t)$ follow inverted fluctuations in $D$ at step $t-1$.
\end{itemize}

\subsection{Short- and long-term consequences of protein designability}
\label{par:designability}

LP structures $S$ differ in how much they are designable, {\em i.e.} in how many sequences $\vb{A}$ have high folding probabilities, say, $p_{nat}(S|\vb{A})>0.99$. Designability has been studied in the literature \cite{li1996emergence,england2003structural}, and it is known that $S_C$ is more designable than $S_A$ \cite{jacquin2016benchmarking}. In practice, to assess the designability of  a given structure, we can either evaluate the entropy $\sigma$ of the Potts model inferred on the MSA (\cref{eq:e_potts}) \cite{barton2016entropy} or compute the mean identity (MId) of the MSA. To do so, we compute the consensus sequence (made of the most frequent amino acids, site after site), and define MId as the average number of sites carrying consensus amino acids. We note that the entropy of the (single-sequence) Potts model is bounded from above by $27 \log 20 \simeq 80.9$, corresponding to a totally unconstrained LP where each amino acid is randomly chosen.

\begin{figure*}
    \centering
    \includegraphics[width=\textwidth]{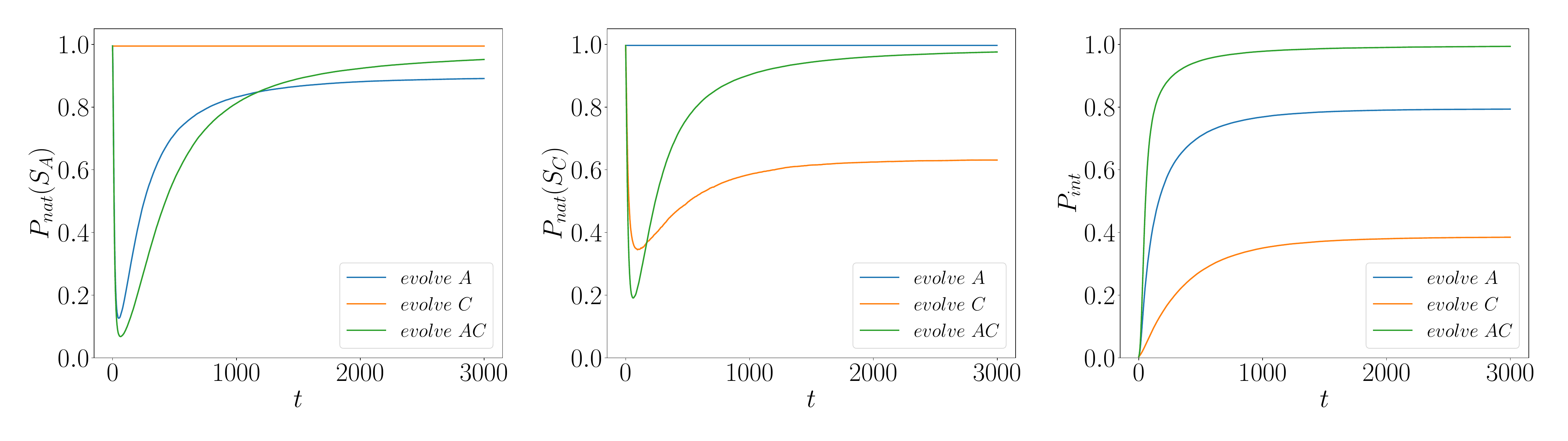}
    \caption{Time dependence of fitnesses $P_{nat}(S_A)$ (left panel), $P_{nat}(S_C)$ (middle panel) and $P_{int}$ (right panel) for the three possible protocols \textit{evolve A} (blue), \textit{evolve C} (orange) and \textit{evolve AC} (green). Same values of stressor, $\bint=20$. MC evolution.}
    \label{fig:different_protocols}
\end{figure*}

The designability of these structures, and how they are affected by the introduction of the binding interaction constraint may help understand the evolutionary trajectories discussed above. We hereafter consider three different classes of binding constraints:
\begin{itemize}
    \item introducing mutations only on $S_A$, keeping fixed $S_C$ (labelled \textit{evolve A}),
    \item introducing mutations only on $S_C$, keeping fixed $S_A$ (labelled \textit{evolve C}),
    \item introducing mutations on both $S_A$ and $S_C$ (labelled \textit{evolve AC}).
\end{itemize}
We report the time dependence of $P_{nat}$ and $P_{int}$ for these three protocols for the same values of $\bnat, \bint$ in \cref{fig:different_protocols};. Protocol \textit{evolve AC} is the most advantageous one, as it produces better configurations in terms of $P_{nat}$ and $P_{int}$. Allowing both sequences to mutate gives rise to a larger number of possibilities to satisfy the constraints; this statement is also confirmed by estimation of the entropy $\sigma ^{fixed} (S_C)\simeq 38.16$ computed on the MSA of {\it evolve C} protocol, and the conditional entropy $\sigma^{cond} (S_C) = \sigma (S_A, S_C) - \sigma (S_A) \simeq 43.55 $ computed on the MSA of {\it evolve AC} protocol.

\begin{figure}
    \centering
    \includegraphics[width=\columnwidth]{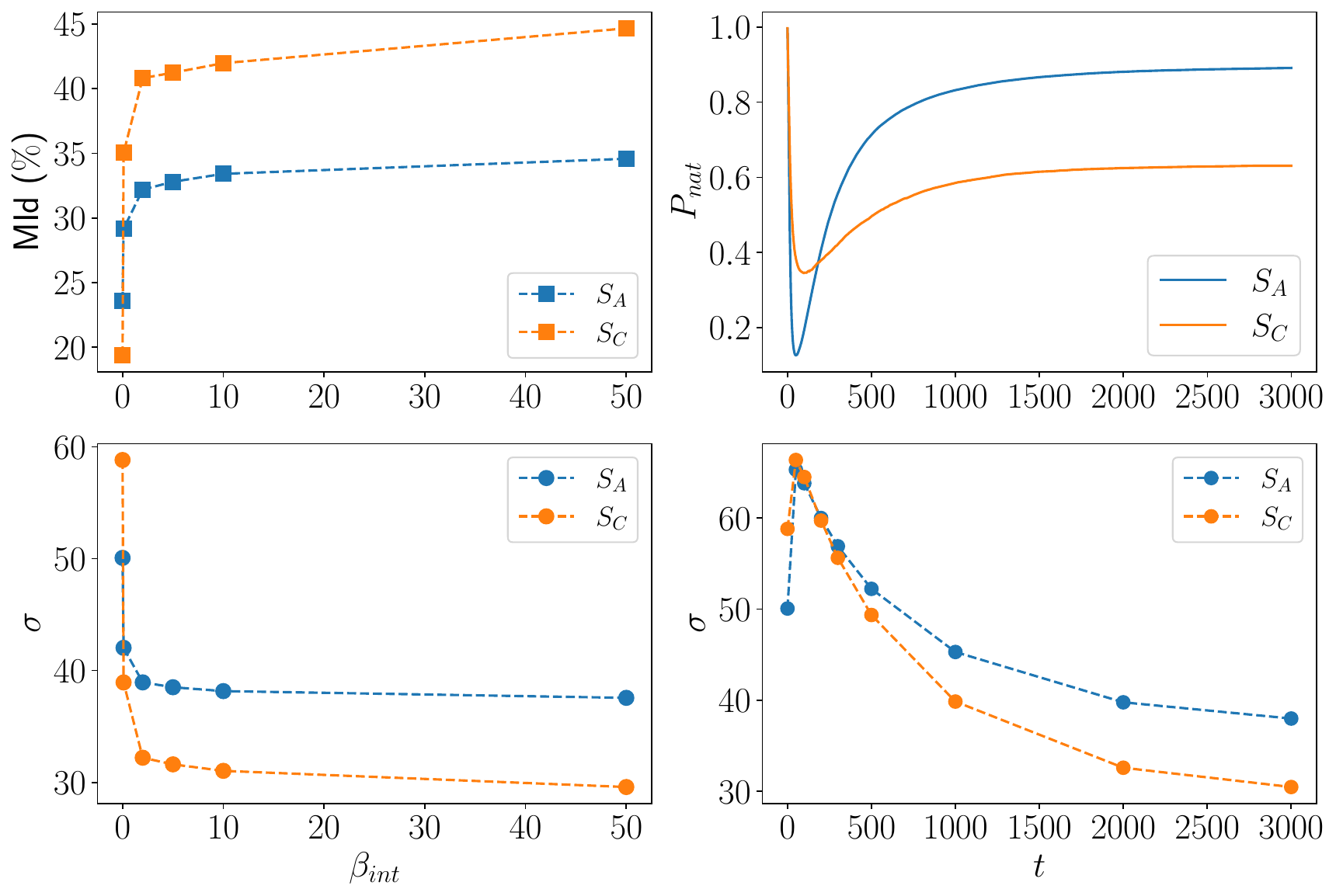}
    \caption{Left: Entropy $\sigma$ (bottom) and mean sequence identity  MId (top) for structures $S_A$ and $S_C$ (blue and orange, respectively) at equilibrium. 
    Right: Time dependence of $P_{nat}(S_A)$, $P_{nat}(S_C)$ (top) and of the entropy $\sigma(S_A)$, $\sigma(S_C)$ (bottom) under protocol \textit{evolve A} and \textit{evolve C}, respectively. Here $\sigma$ at intermediate time is computed from the Potts model inferred on the MSA at that time. Same values of stressor, $\bint=20$. MC evolution.}
    \label{fig:pnat_entropy_designability}
\end{figure}

We report in \cref{fig:pnat_entropy_designability} (left panel) the entropy $\sigma$ together with the MId for several values of $\bint$ and for $S_A$ and $S_C$, evolved with, respectively, protocols \textit{evolve A} and \textit{evolve C}. The reported values are relative to end-point evolution, \ie{} at equilibrium. As the stressor intensity $\bint$ increases, the designability of both structures decrease, as it is harder for sequences to cope with the constraint on $P_{int}$. For all tested $\bint > 0$, the structure $S_A$ is realized by more sequences than $S_C$, contrary to what happens for non-interacting structure ($\bint=0$). 

We again resort to the Potts model inferred from sequence data at different time points during evolution to characterize this phenomenon. In \cref{fig:pnat_entropy_designability} (top right panel) we observe that the $P_{nat}$ cross each other, signalling  an inversion in the designability of the two structures, see \cref{fig:pnat_entropy_designability} (bottom right panel). In other words, evolving a dimer surface through binding to a fixed protein is more harmful in terms of fitness when the fixed protein has low designability. It is worth mentioning that this result is valid in general and does not depend on the particular choice of the two structures, as we validate in \cref{sec:addition_dimer} using different structures.  

\subsection{Microscopic mechanisms}

\subsubsection{Propagation of constraints on the interacting face}

\begin{figure}
    \centering
    \includegraphics[width=\columnwidth]{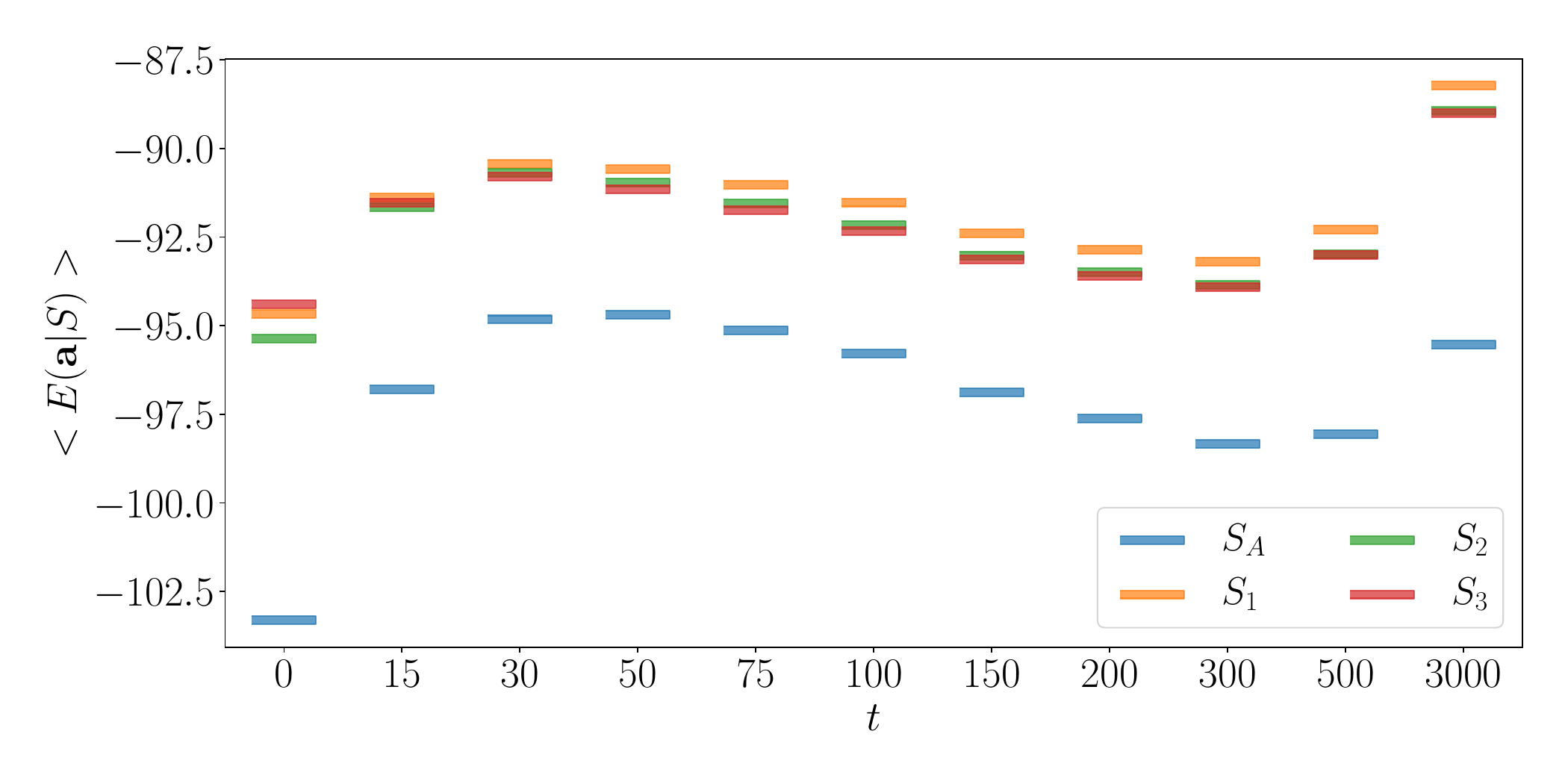}
    \caption{\label{fig:energy_levels}Time evolution of the mean energy  for protein sequences folded in $S_A$ and in its first three competing structures (labelled as $S_1$, $S_2$ and $S_3$, see Appendix~\ref{par:competing}), namely the three structures with energies closest to $S_A$. All other structures have energy levels higher than $S_3$. We note that the three competing structures share with the native folding $S_A$ more than $10$ contacts, whereas an average random structure share just $5$ contacts (see \cref{par:competing}). Interestingly, there is an inversion in the order of competing structures going from $t=0$ until the equilibrium state at $t=3,000$.}
\end{figure}

\begin{figure*}
    \centering
    \includegraphics[width=.9\textwidth]{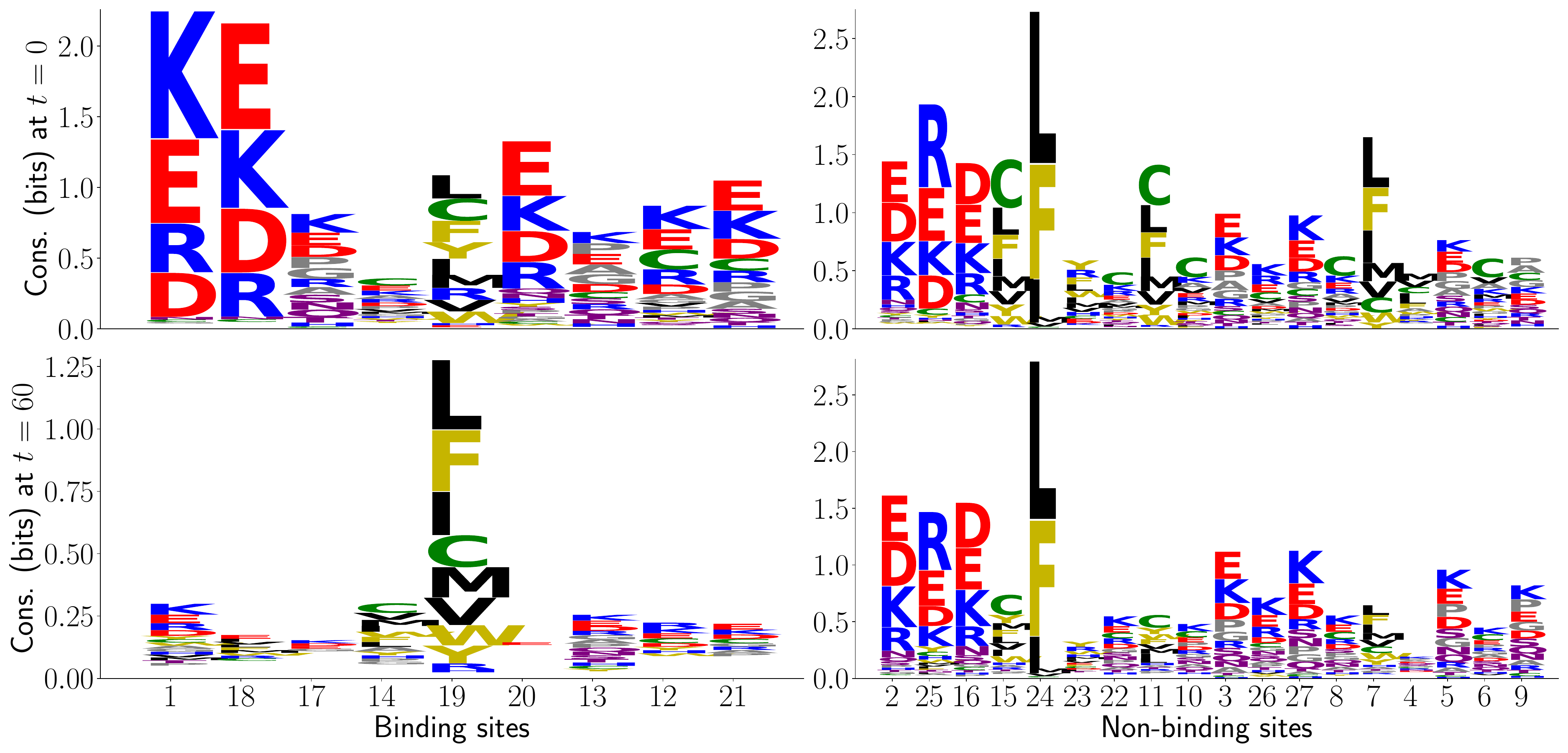}
    \caption{\label{fig:seq_logos}Sequence logos showing amino acids conservation on each site for protein sequences folded in $S_A$, averaged over the full MSA. On each site, the total height corresponds to the local conservation (see~[18]) and the letter sizes correspond to their site-frequencies: the bigger a letter is, the more frequent that amino acid is in the MSA. We compare sites involved in binding (left) and non-binding sites (right) at $t=0$ (top) and $t=60$ (bottom) to show larger mutability on sites in interaction with the other protein. Sites $1-18$ (or, equivalently, $2-25$) form a crucial electrostatic mode in the native structure (top row), which is lost in correspondence to the minimum of $P_{nat}$ (bottom row). Color code: blue for basic, red for acidic, green for Cystein, black for hydrophobic and gold for aromatic ones.}
\end{figure*}

\begin{figure}
    \centering
    \includegraphics[width=\columnwidth]{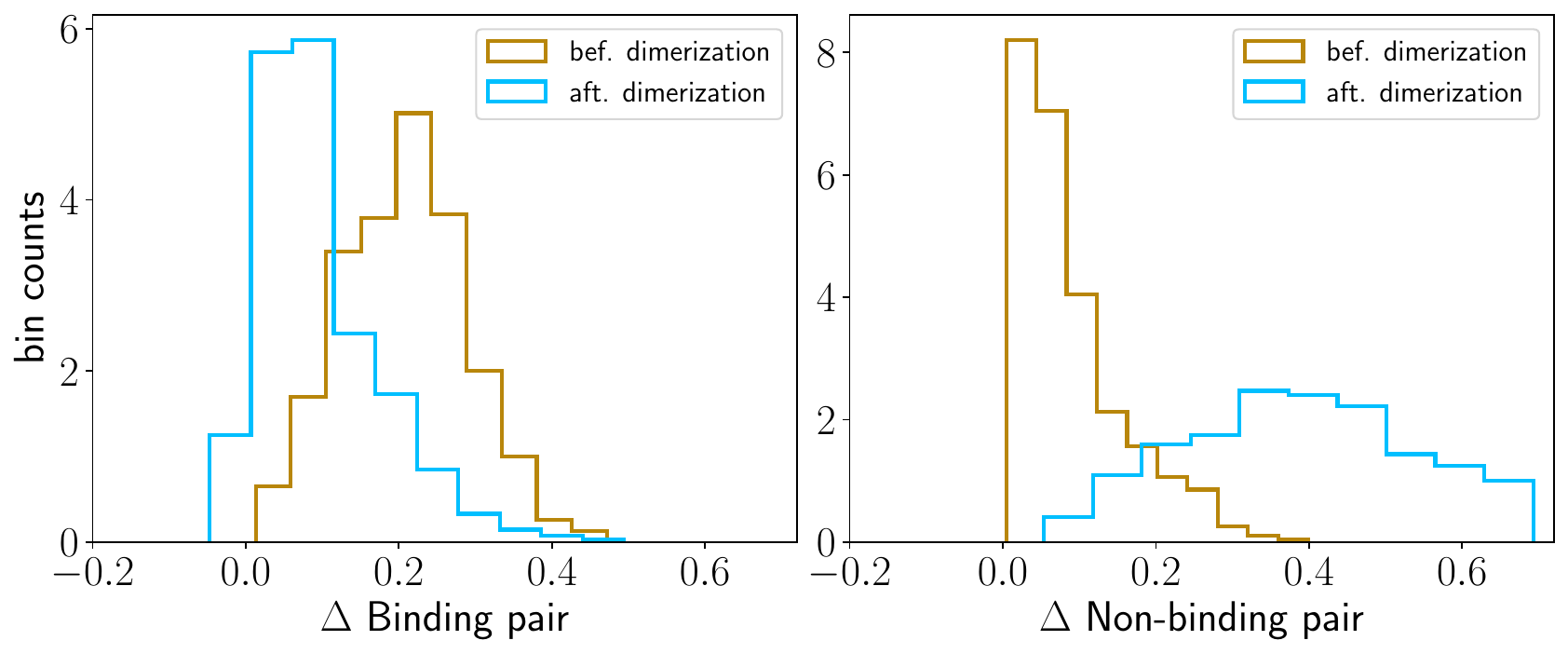}
    \caption{\label{fig:res_frust}Distribution of scores $\Delta$ for two contact pairs, one involved in binding (left, pair $1-18$ in \cref{fig:dimer}) and the other not involved (right, pair $3-26$ in \cref{fig:dimer}), before and after dimerization takes place (gold and sky blue histograms, respectively). The scores are computed over the MSA of structure $S_A$ at $\bint=5.0$. Similar results hold for other contact pairs.}
\end{figure}

\begin{figure}
    \centering
    \includegraphics[width=\columnwidth]{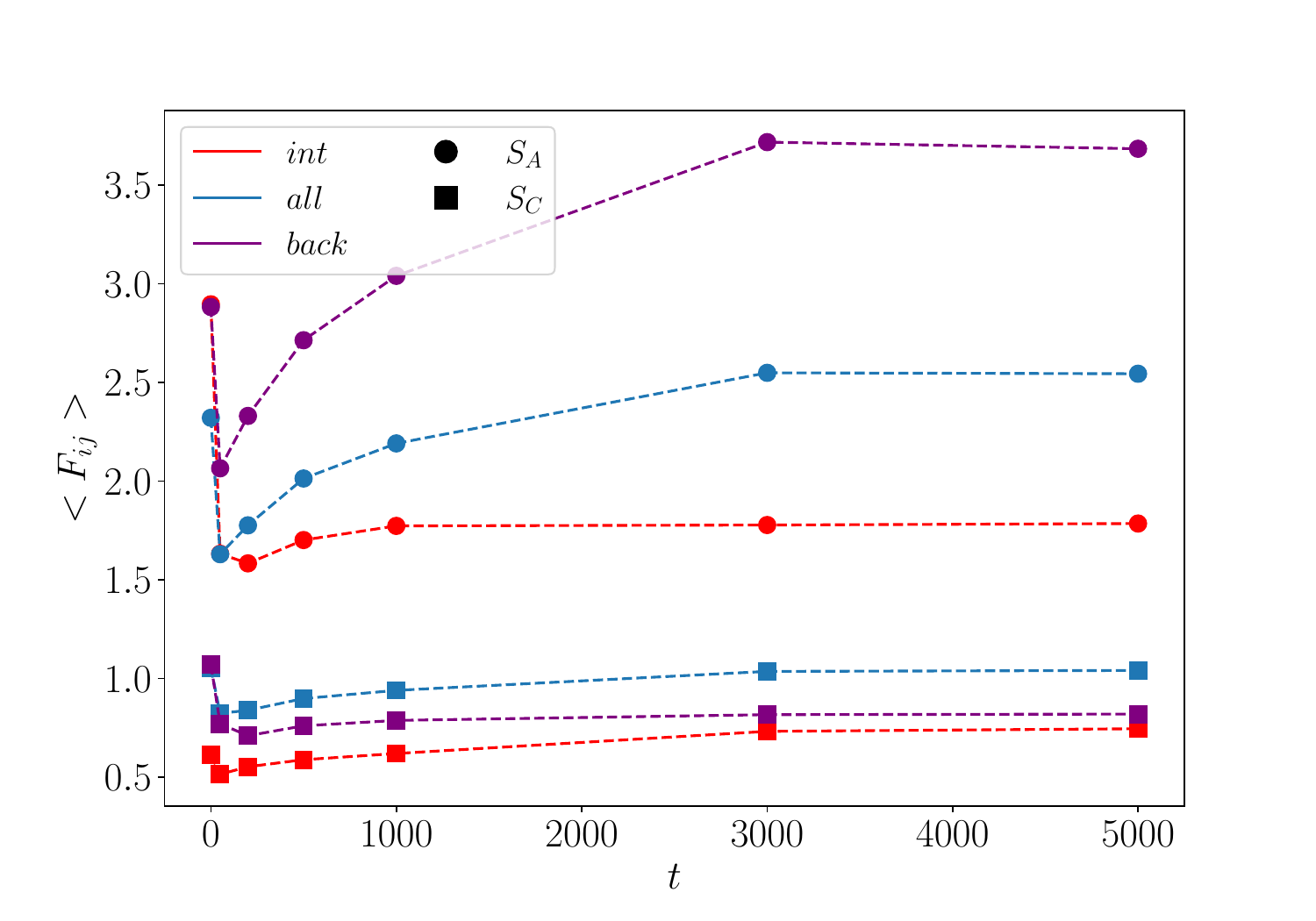}
    \caption{\label{fig:FB-norm}Time evolution of $\expval{F_{ij}}$ averaged over contacts on interacting face (red), on back faces (purple) and over all $28$ contacts (grey), for structure $S_A$ (circle) and $S_C$ (square). See color code of \cref{fig:dimer} for red and purple faces we are referring to. The norm is computed for a MSA at $\bint = 2.0$.}
\end{figure}

The consequences of having a binding selection pressure $\beta_{int}$ during the evolution can  be seen form the point of view of single protein structure and folding stability. At the beginning of the evolution ($t=0$), each fold is very stable, which corresponds to large energy gap between the native structures and the competing structures (cfr. \cref{fig:energy_levels} and \cref{par:competing}). As the evolution starts both proteins become less stable for some time, as the energy levels of competing structures  get closer to the native folds. Eventually, as equilibrium is approached and the dimer is formed, a large energy gap is restored (albeit smaller than at $t=0$). \cref{fig:energy_levels} shows the energy levels for the first competing structures of $S_A$ during a typical MC evolution. 
The minimal energy gap is reached at similar evolutionary time points, regardless of the stressor strength $\beta_{int}$. This time corresponds to sequences where the amino acids on the interacting face have been mutated to favor the binding mode, thus destabilizing the native fold and making alternative competing folds more likely. Statistically, the time needed to update - at least once - all the nine amino acids on the interacting face is $\tau_{typical} \sim 9 / p$, where $p$ is the probability of proposing and accepting a mutation on a site of the interacting face. We estimate it from the MC simulation as the acceptance rate of mutations on interacting sites. It results $p \sim 0.12$ \footnote{This estimate is based on the short-time transient, while the overall acceptance rate is much lower.}, giving $\tau_{typical}$ in agreement with the one observed in \cref{fig:evolution}. 
The sequence logos \footnote{ \label{tab:seqlogo}Sequence logos consist of a visual representation of conserved amino acids expressed in bits. On the $i$-th site the total height is $C_i = \log_2 Q - \sum_{a=1}^{Q} log_2 f_i(a) $, where $Q$ corresponds to the total number of amino acids; the size of each letter is then proportional to its frequency $f_i (a)$. }, displayed in \cref{fig:seq_logos} for $S_A$, visually show enhanced mutability on interacting face compared to other faces during the highly unstable transient. For example, positive (blue) and negative (red) charged residues on binding sites $1, 18$ are highly conserved in the native structure, as they form an electrostatic bridge with amino acids of opposite charge. Non-binding sites $2,25$ also display an electrostatic mode in the native structure. At $t=60$, the  electrostatic contact $1-18$ is depleted due to the ongoing binding with the other structure, while contact $2-25$ remains less impacted. Enhanced mutability allowing for dimerization leads to residual frustration at single monomer level, which can be quantitatively evaluated along \cite{ferreiro2007localizing}. Given a contact pair $i-j$, for each sequence $\vb{A}$ in the MSA we compute $P_{nat}(S| \vb{A})$ and $P_{nat}(S | \vb{A'})$, where $\vb{A'}$ differs from $\vb{A}$ by replacing amino acids $a_i, a_j$ with two other amino acids chosen uniformly at random. We then define the following score 
\begin{equation}
    \Delta_{i,j}(\vb{A}) = P_{nat}(S| \vb{A}) - \left< P_{nat}(S | \vb{A'}) \right>, 
\end{equation}
where the average is carried out over all possible pairs of amino acids on $i,j$ excluding the actual one. The distribution of scores across the MSA of structure $S_A$ is shown in \cref{fig:res_frust} for a contact pair involved  (\ie{} binding pair) or not (\ie{} non-binding pair) in the binding of the dimer. Upon dimerization, the scores of binding pairs are shifted to lower values: those pairs of sites now carry amino acids mostly optimized for binding, and almost any other amino-acid pair result in the same $P_{nat}$ value. Therefore, such pairs are not crucial to maintain structural stability and we deem them as highly frustrated. Residual frustration at single monomer level is again the trademark of stability-affinity competition.    \\
The presence of evolutionary trade-offs between $P_{nat}$ and $P_{int}$ also affects contacts' strength, assessed by the  Frobenius norm computed as described in \cref{par:methods}. In \cref{fig:FB-norm} we show the evolution of $F_{ij}$ during the MC simulation, for both structures. For the LP folded in $S_A$ the contacts belonging to the interacting  face are the ones that suffer most from the binding constraint, as the corresponding Frobenius norms undergo large drop; conversely, the Frobenius norms associated to the back face (opposed to the interacting face) reach larger values after equilibration than the initial time point. \\
The Frobenius norm of structure $S_C$ is smaller at any time point, suggesting that there is less coevolutionary pressure between sites due to the native design in structure $S_C$ compared to $S_A$ (cf. \cref{fig:FB-norm}); this is consistent with structure $S_C$ being more designable than $S_A$. 

\subsubsection{Learning the binding mode through local fields}

In this Section we want to better characterize the selection pressure imposed by the binding mode through $\bint\ne 0$. We use the {\it evolve A} protocol and model the binding interactions as external fields $h_i$ (exerted by the protein $S_C$) on the sites of the structure $S_A$ belonging to the interacting face. To check whether this simple modeling approach, which neglects couplings $J_{ij}$ within the interacting face is plausible we develop a slightly {\it modified protocol} of MC evolution, where at each Metropolis step, $P_{int}$ in 
\cref{eq:pint} is replaced with
\begin{equation}
\label{eq:pint_fields}
    P^M_{int}(S_1 + S_2 | \vb{A_1}, \vb{A_2}) \propto e^ {- \mathcal{E}_{int} (\vb{A_1},\vb{A_2} | S_1 + S_2, 1 )}.
\end{equation}
In practice, discarding the denominator in \cref{eq:pint} amounts to approximate $P_{int}$ as a product over the sites of the binding interface. The competition between interfaces, usually leading to negative design, is therefore neglected.

This {\it modified framework} gives a qualitatively similar behaviour for the time dependencies of $P_{nat} (S_A)$ and of $P_{int}$. We learn a Potts model from the sequence data in the same way as \cref{par:dimer_characterization}, expecting that the inferred fields of sites on the interacting faces are highly correlated with the Myazawa-Jernigan (MJ) energy matrix. This is especially true when the stability of the protein is  compromised, \ie{} in the low $P_{nat}$ region, because the binding selection pressure (fields) dominates over the internal contacts (couplings). We show in Table~\ref{tab:Corr_coeff} the average correlation coefficients between inferred fields and the MJ matrix for sites on the interacting face and not, at two different MC time steps. 

\begin{figure}
    \centering
    \includegraphics[width=\columnwidth]{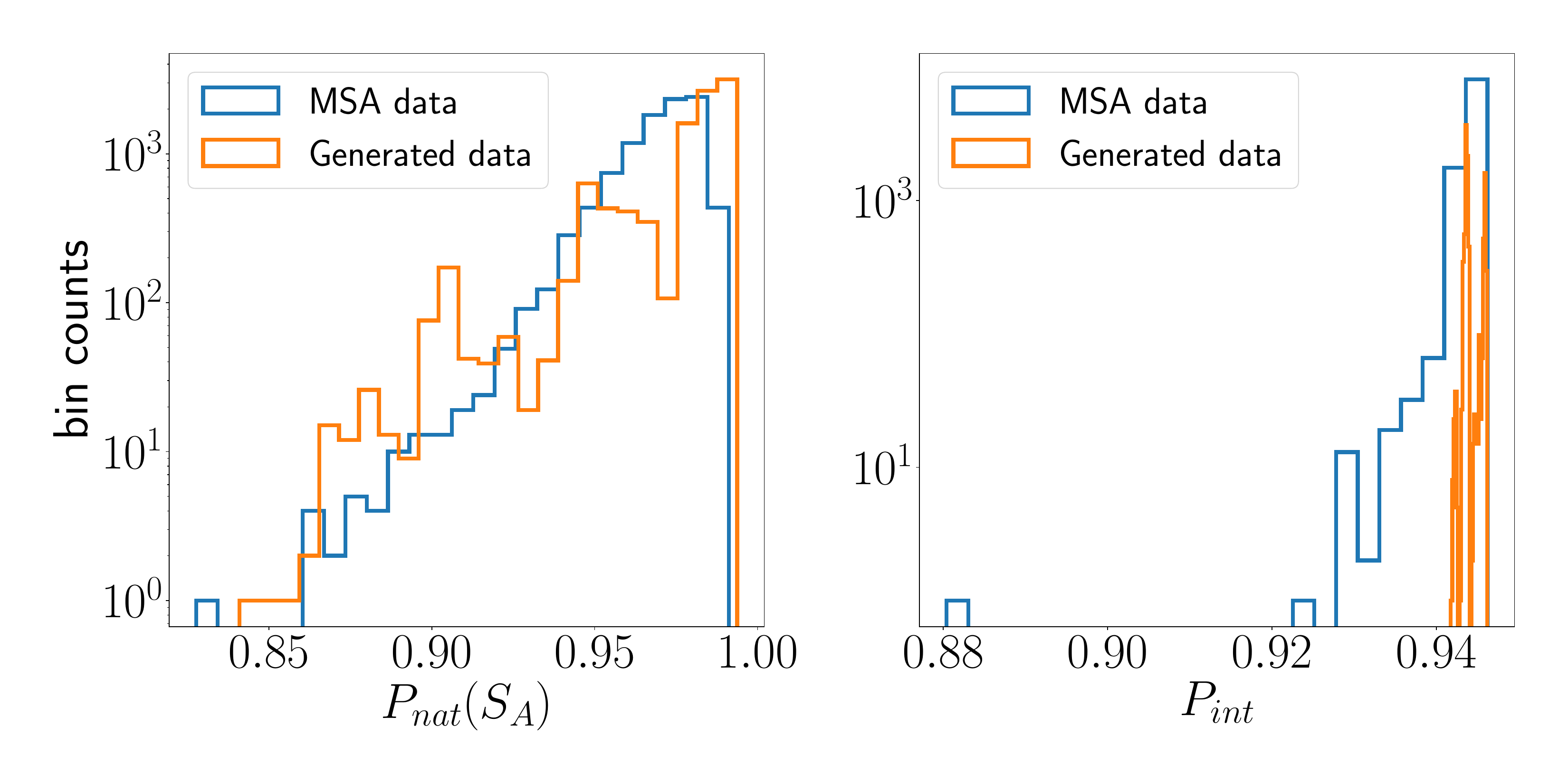}
    \caption{Distribution of $P_{nat}(S_A)$ (left) and $P_{int}$ (right) for the training dataset and the generated MSA. To ensure high fitnesses on the new sequences, we sample the space with low temperature, $T=0.2$.}
    \label{fig:BM_fields_generation}
\end{figure}

To assess the validity of this field-based model, we consider a new learning procedure for the Potts model that consist in two steps. We collect in our dataset sequences having $S_A$ as their native conformation ($P_{nat}(S_A) > 0.99$), and produce a smaller dataset with sequences for $S_A$ all bounded to the same sequence folded in $S_C$. Inference of the Potts model works as follows:
\begin{enumerate}[(i)]
    \item we learn couplings and fields of the Potts model on the first dataset, having high $P_{nat}(S_A)$ and  $P_{int}\simeq 0$;
    \item we now learn fields of the Potts model on the second dataset,  using less data ($25\%$ of sequences used for the previous step), and with couplings $J$ frozen to their values obtained in step (i). 
\end{enumerate}
In other words, we first have a background Potts model, able to model the distribution of $S_A$ sequences, and then we infer only new fields to capture the features related to the interaction with the other protein sequence. We show in \cref{fig:BM_fields_generation} that learning local fields is enough to generate good sequences in structure $S_A$ that bind well with the given protein sequence in $S_C$. Inferred local fields are sufficient to reproduce the $P_{nat}$ distribution (see \cref{fig:BM_fields_generation}, left panel), suggesting that couplings $J_{ij}$ inferred at initial stage with no interaction are still meaningful. However, if one removes couplings at the single structure level attempting to use only fields to reproduce the bounded dimer distribution, the approach completely fails: it means that only inter-protein interactions can be modelled with local fields. 
Let us note that the generated sequences have mean identity (as defined in \cref{par:designability}) MId $\sim 80 \%$, compared to $60 \%$ for the training data, likely due to the presence of strong fields that force interacting sites to be very conserved. The generated sequences have an average Hamming distance of $11.75$ amino acids to the ones used in the training.

\begin{table}
\begin{tabular}{ |p{1.5cm}|p{2.5cm}|p{3.5cm}|  }
 \hline
  & Interacting sites &Non-interacting sites \\
 \hline
 $t=0$   &  $0.05$   & $-0.08$ \\
  \hline
 $t=100$&   $0.92$  & $-0.09$ \\
 \hline
\end{tabular}
\caption{Correlation coefficients of BM fields versus the MJ energy $\expval{E(a_i,a_j)}_{a_j}$ averaged over columns, for the model inferred on the {\it modified case} scenario. At $t=0$ both fields of interacting and non-interacting sites are not (or negatively) correlated to the MJ values; after evolving the dimer surface, a strong positive correlation can be observed for sites on the interacting face.}
\label{tab:Corr_coeff}
\end{table}

\section{Discussion}
\label{sec:conclusion}
In this work we analyze the case of two lattice proteins that evolve a dimer surface under a binding selective pressure, through a Monte Carlo Metropolis or Population Dynamics approach. 
In \cref{sec:Equilibrium_prop} we study the equilibrium properties for different stressor strengths $\bint$, using the inference of a Potts model over the dimer MSA to reveal the effect of foldability-dimerization trade-offs on inter- and intra-protein couplings.   
In particular, we reproduce the majority of internal contacts of both structures, with contacts in $S_C$ always less predictable than in $S_A$ due to the higher designability of the former structure. As for the binding contacts, we succeed in reconstructing eight out of nine of them, missing always the central contact. In fact, the latter is shared in the functional and non-functional rotated modes, and our inferred model is not able to capture the \textit{negative design} associated to such contact of competing modes \cite{jacquin2016benchmarking}. We also characterize the dimer interaction, assessing the quality of the function binding mode over the competing ones. \\
We then study in \cref{par:evolutionary_tradeoff} the ongoing evolutionary trade-offs during the dimer formation. Putting a selective pressure $\bint$ forces the two LPs to explore unstable conformations in order to maximize the binding, at the expense of folding; this results in a temporary loss of stability, which depends on how rapidly the stressor $\bint$ is applied during sequence evolution and how strong it is. Indeed, the drop in the native fitnesses $P_{nat}$, has a strong out-of-equilibrium nature; whereas applying a smooth selection pressure relaxes the constraint over the two interacting faces and allows for a monotonically (decreasing) evolution of $P_{nat}$. This observation signals the existence of a minimal time for adaptation to new constraints, in agreement with experimental findings for bacterial evolution under stressful conditions \cite{iwasawa2022analysis,yen2017history,swings2017adaptive}. \\
We then resort to the evolution of a non-clonal population (see \cref{par:non_clonal_pop}) to understand the interplay between the population size, the mutation rate per individual and the stressor strength. We see that population size here acts as an inverse temperature that sets the stringency of the fitnesses in evolution of stable complexes, which accounts for the lower values of $\bnat$, $\bint$ used in the Population Dynamics compared to MC algorithms. This statement is especially clear when dealing with mono-clonal evolution, where the population size and the fixation probability of a mutation are related through \cite{kimura1962probability}
\begin{equation}
    P_{fix} = \dfrac{1 - \exp (-2s)}{1 - \exp(-2Ns)},
\end{equation}
where $s$ is the selection coefficient and $N$ the population size. Hence, here $N$ is playing the role of an inverse temperature (see \eg{} \cite{sella2005application,rotem2018evolution}).
Furthermore, we discuss how the selection pressure applied on the dimer evolution has consequences on the designability of the structures both at the short- and long-term level. We show that high stressor strengths reduce the diversity of MSA and that binding to a more designable folding (cf. \textit{evolve A} protocol) allows to find more optimal sequences. \\
Eventually, we discuss the microscopic mechanisms underlying the dimer formation showing that the binding interaction can be efficiently encoded in a Potts model on the single structure with local fields that mimic the selective pressure arising from the other structure. \medskip

In the future we plan to apply the discussed framework to real data where one can experience evolutionary trade-offs, e.g. bacteria strains evolving under two (or more) competing stressors.

\vskip .3cm
\noindent {\bf Acknowledgements:}
E.L. is funded by the CNRS - University of Tokyo \textit{"80 Prime"} Joint Research Program. 

\appendix
\label{par:methods}
\section{Inferring Potts model}
We use a Boltzmann Machine (BM) to learn the probability distribution of the data and thus inferring the parameters of the Potts model. A BM is a probabilistic graphical model constituted of a single set of random variables $\vb{v} = (v_1 , \dots ,v_N)$ that interact within each other through a coupling matrix $\vb{J}$ and that are subject to local fields, hereafter called $g_i$. For our purpose, we need a set of $N=54$ variables corresponding to the length of the two adjacent amino acids sequences, with each variable assuming $q=20$ different states. For such BM the probability distribution of the variables set $\vb{v}$ is exactly given by the Gibbs-Boltzmann probability at fixed temperature $T_{inf} =1$ with energy 
\begin{equation}
    E(\vb{v}) = - \sum_i g_i(v_i) - \sum_{i<j} J_{ij} (v_i, v_j),
\end{equation}
where the couplings $J_{ij}$ and fields $g_i$ set the mean and correlation of the  variables $v_i$.
Training a BM to infer its parameters consists in fitting numerically the distribution $P$ of the data by maximizing the likelihood $\mathcal{L} = \expval{\log P}_{data}$. Taking the gradient of $\mathcal{L}$ and setting it to zero for its maximization, bring us to solve the following problem (for a generic parameter $\theta$ of the model)
\begin{equation}
\label{eq:gradient_problem}
    \nabla_{\theta} \mathcal{L} = - \expval{\nabla_u E(\vb{v})}_{data} + \expval{\nabla_\theta E(\vb{v})}_{model}
\end{equation}
where $\expval{\cdot}_{data}$ stands for the expectation value over the data, while $\expval{\cdot}_m$ over the model. The gradient update thus consists of decreasing the energy of the data configurations while increasing the ones of the model. In our case, the gradient problem in \cref{eq:gradient_problem} can be turned into the set of equations
\begin{equation}
\label{eq:gradient_loglikelihood}
    \begin{split}
        \pdv{\mathcal{L}}{g_i} &= \expval{v_i}_d - \expval{v_i}_m \\
        \pdv{\mathcal{L}}{J_{ij}} &= \expval{v_i v_j}_d - \expval{v_i v_j}_m,
    \end{split}
\end{equation}
which is a momentum-matching problem. The fitting procedure stops when the two expectation values match. Among the two right hand terms in \cref{eq:gradient_loglikelihood}, it is easy to compute the average over the data as it can be done from the dimer MSA just once when the learning procedure starts; on the other hand computing the average over the model is a challenging task. Here, we use standard Gradient Descent (GD) with a varying learning rate in time and we make use of Persistent Contrastive Divergence (PCD) algorithm with a fixed number of MC steps between each update for sampling data to compute the average over the model. To avoid over-fitting during training, we use a $L_1 ^2$ regularization term with strength $\lambda_1^2$. \\
All in all, we tune the hyper-parameters for learning as follows
\begin{itemize}
    \item Number of epochs = 150 
    \item Learning rate = 0.005 (it decays with a 0.5 rate after half iterations)
    \item MC steps between each update = 5
    \item $\lambda_1^2$ = 0.025
\end{itemize}
\medskip
Given the BM parameters, we can also compute the entropy $\sigma$ of the model as the opposite of the log-likelihood averaged over the full MSA, \ie{}
\begin{equation}
    \sigma = - \expval{\log P (\vb{v})}_{\vb{v}},
\end{equation}
which involves the numerical computation of the partition function $\mathcal{Z}$. The latter is intractable as it consists in summing over all configurations $\vb{v}$; therefore, we estimate it with the Annealed Importance Sampling (AIS) algorithm \cite{neal2001annealed}.

\section{Dimer sequence generation}
\begin{figure}
\centering
    \includegraphics[width=\columnwidth]{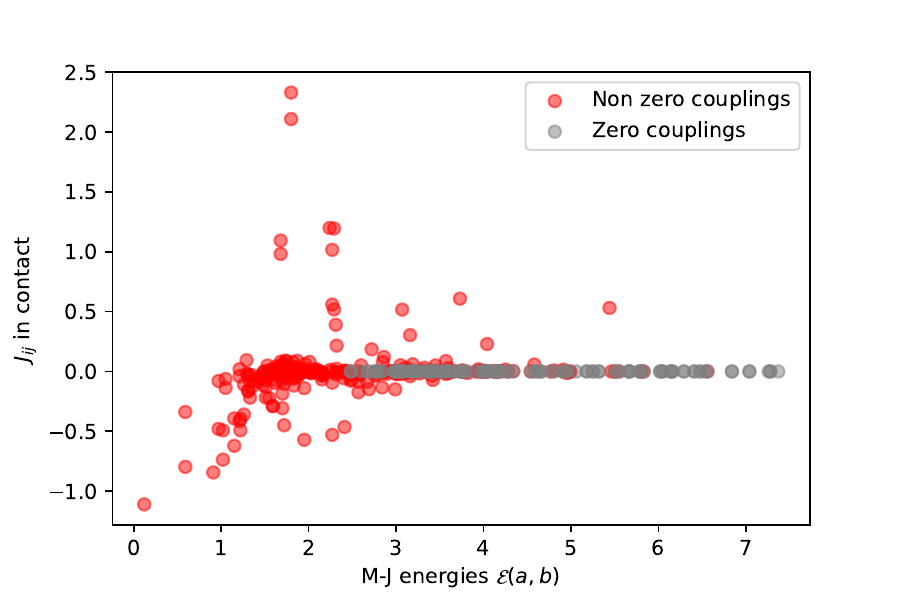} 
    \caption{\label{fig:Jij_contactsite}Couplings matrix $J_{ij} (a, b)$ for two specific sites in contact, \ie{} $ i = 3, j = 25$ of structure $S_A$, versus the MJ energy matrix $\mathcal{E} (a,b)$.}
\end{figure}

We generate new dimer sequences using the BM log-likelihood as the energy in the Gibbs sampling. For computational reasons after a long thermalization, we run $100$ chains in parallel each of length $1,000$, and sample a dimer sequence each $500$ steps. Sampling with the inferred Potts model is hard because we must have reconstructed the $J_{ij}$ matrix very precisely; however the BM has learnt what are the sites in contact but it is struggling to understand which couple of amino acids is present on a given pair of sites. Effectively, if we look at the matrix $J_{ij}$ for a given pair of sites in contact, most of its entry are zeros (\ie{} the BM has never seen such pair of amino acids in the MSA), some are slightly different from zero and few of them have large values (\ie{} those associated to polar amino acids). We show this trend in \cref{fig:Jij_contactsite}. Thus with so few peaks in the matrix, it is hard to generate and to ensure a great diversity of the sampling space, because the BM is exploring just one deep minima in the landscape. A solution for that would be to enlarge the effective depth of the MSA with more diverse dimer sequences, \ie{} sampling with MC at higher temperature.

\section{Additional dimeric assemblies}
\label{sec:addition_dimer}
\begin{figure}[h!]
    \includegraphics[width=\columnwidth]{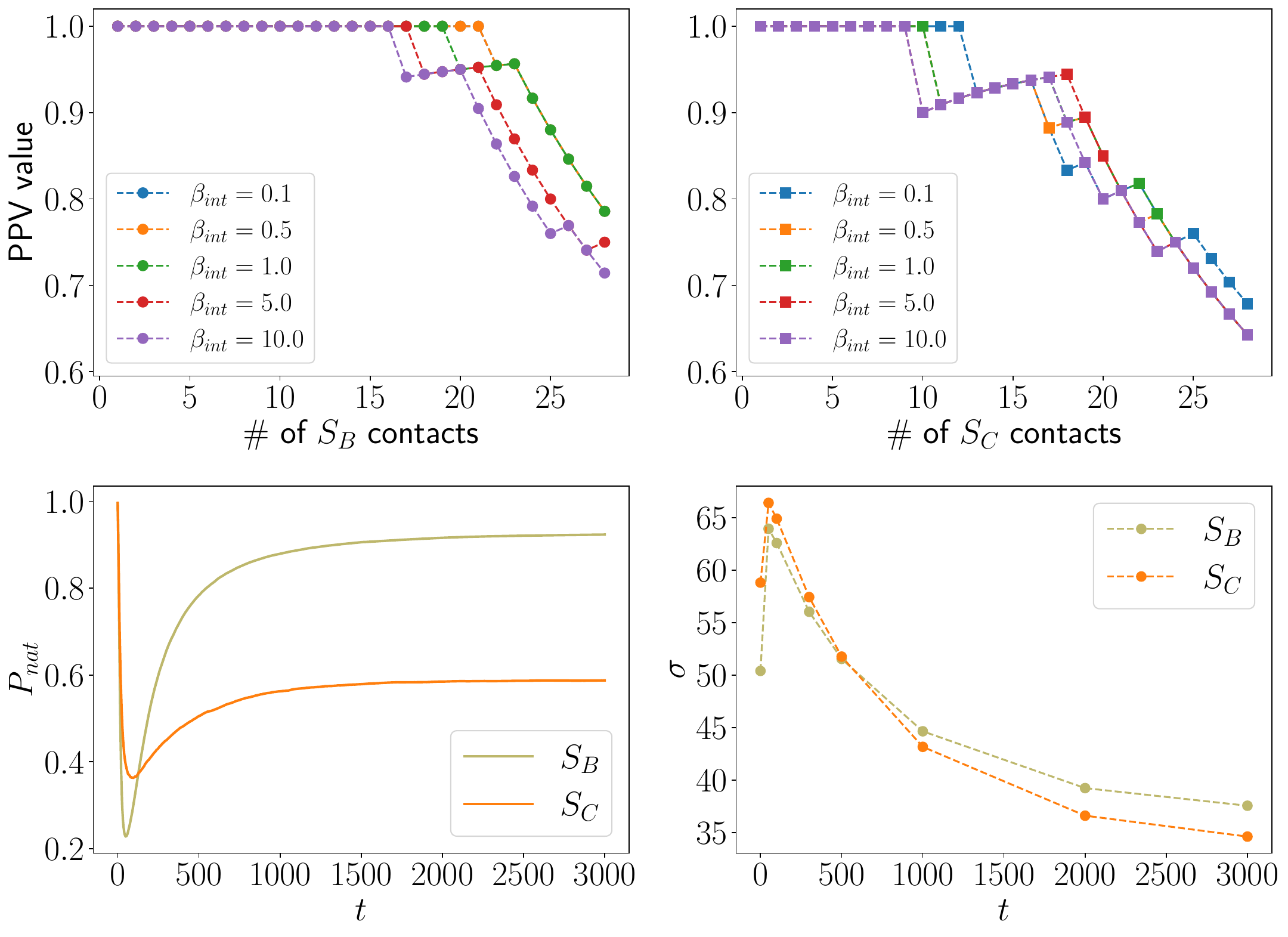}
    \caption{Top: PPV for structure $S_B$ (top left panel, circles) and structure $S_C$ (top right panel, squares) at different values of $\bint$. The MSA is made of $15,000$ dimer sequences evolved for $3,000$ MC steps. Bottom: Time dependence of $P_{nat}(S_B)$, $P_{nat}(S_C)$ (bottom left panel) and entropy $\sigma(S_B)$, $\sigma(S_C)$ (bottom right panel) under protocol \textit{evolve B} and \textit{evolve C}, respectively. Same values of stressor, $\bint=20$. MC evolution.}
    \label{fig:BC_dimer_SI}
\end{figure}

To further corroborate our results we perform an equivalent analysis using a different LP monomer, namely $S_B$ as labelled  in \cite{jacquin2016benchmarking}, that binds $S_C$. Binding of new different native structures yield the same scenario, suggesting that our results presented in the main text go beyond the specific conformation of $S_A$, $S_C$. Here, to speed up computation we restrict ourselves to $15,000$ MSA. We show results for the dimer $S_B - S_C$ in \cref{fig:BC_dimer_SI}, where we plot the PPV for intra-structure contacts as we do in \cref{fig:ppv} for the dimer $S_A - S_C$. Since $S_B$ is less designable than $S_C$, the picture here confirms that high designability negatively affects contact predictions of the native structure. As in \cref{fig:ppv}, the worst predicted couplings involve the central site of the native structure and/or the central site of the binding face. \\
Furthermore, we use structure $S_B$ in interaction with $S_C$ to validate our claim that evolution of a dimer keeping one of the two interface fixed is more or less harmful in terms of fitness depending on the designability of the fixed interface (see \cref{par:designability}). We thus designed, as we did for the dimer $S_A - S_C$, the two protocols \textit{evolve B}, \textit{evolve C} and computed the entropy at different time steps for both protocols. As in \cref{fig:different_protocols}, we can see in \cref{fig:BC_dimer_SI} that the fitness $P_{nat}$ is lower along protocol \textit{evolve C}, where we keep fixed protein $S_B$ that is less designable than $S_C$.

\section{Competing structures}
\label{par:competing}
\begin{figure}[h!]
\includegraphics[width=\columnwidth]{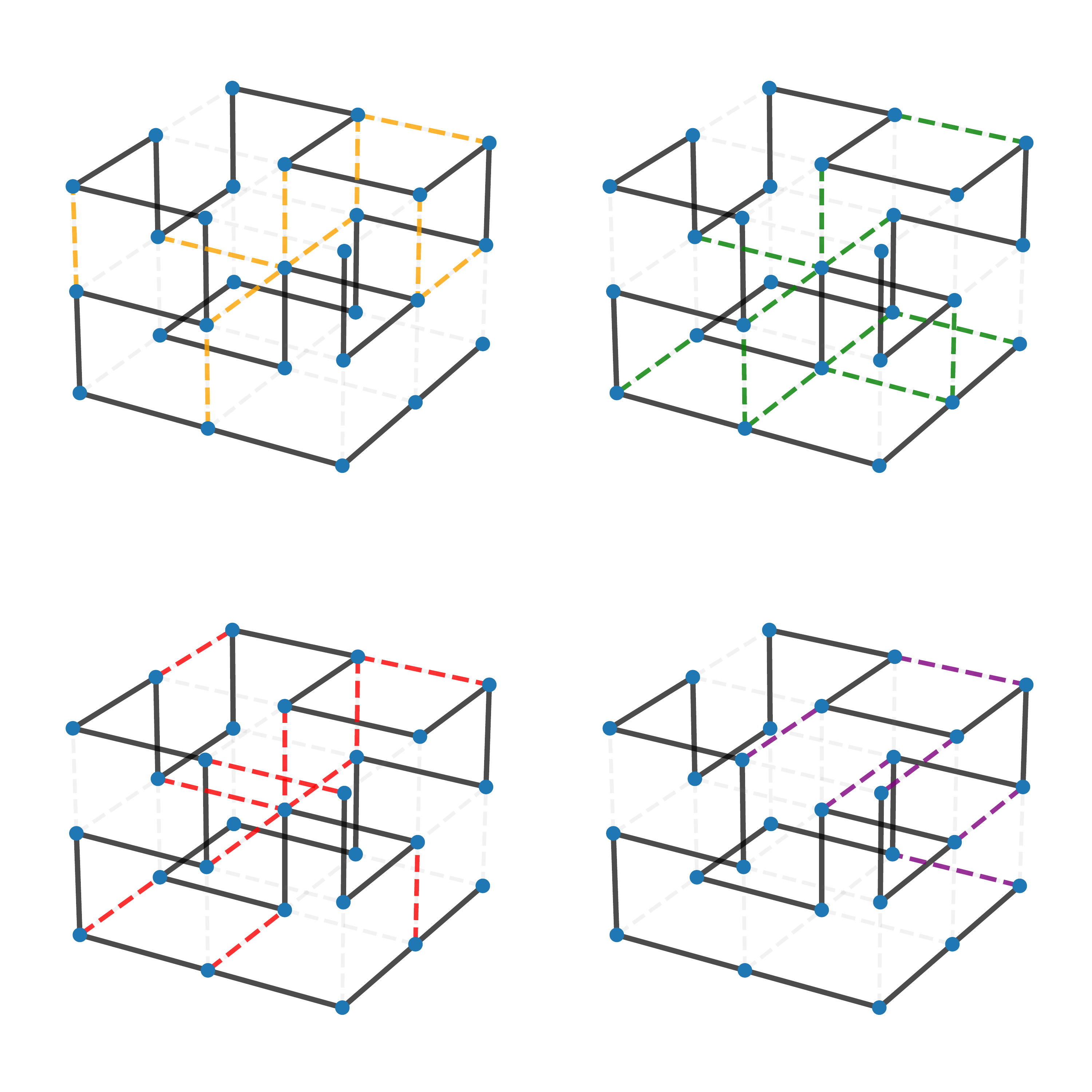}
\put(-225,127.5){$\mathbf{S_A}-\mathbf{S_1}$   \qquad \qquad  \qquad \qquad \qquad $\mathbf{S_A}-\mathbf{S_2}$}
\put(-225,7.5){$\mathbf{S_A}-\mathbf{S_3}$   \qquad \qquad  \qquad \qquad \qquad $\mathbf{S_A}-\mathbf{S_r}$}
\caption{\label{fig:comp_foldings}Representation of the competing foldings of $S_A$, as discussed in \cref{fig:evolution}. Solid black line represents the backbone of structure $S_A$. Dotted grey shaded lines highlight contacts only present in the native folding $S_A$. Colored dotted lines show contacts of $S_A$ common to its competing structures (orange, green and red refers to $S_1$,$S_2$,$S_3$, respectively); purple dotted lines show contacts of $S_A$ shared with a random structure $S_r$.}
\end{figure}

In \cref{fig:comp_foldings} we report the folding of the first three competing structure with $S_A$, labelled in the main text as $S_1$, $S_2$, $S_3$ in \cref{fig:evolution}.
Such structures have been identified among the $\mathcal{N}=10,000$ representative foldings, as the ones having the smallest energy gap with the native structure. The energy for a given folding $S$ has been measured as in \cref{eq:E_MJ}, and averaged over $46,000$ sequences that fold in $S_A$. The random structure $S_r$ has been randomly selected and has a large energy gap with the native structure. The more contacts structure $S$ shares with the native folding, the more such structure will be considered as a competing one. 

\newpage
\nocite{*}
\bibliography{references}

\end{document}